\documentclass[letterpaper,twocolumn,10pt]{article}
\usepackage{epsfig, endnotes, amsmath, amsfonts, amssymb}
\usepackage{usenix2019_v3}
\usepackage{bm}
\usepackage{dsfont}
\usepackage{microtype}
\usepackage{siunitx}
\usepackage{tikz}
\usepackage{url}
\usepackage{graphicx}
\usepackage{booktabs}
\usepackage{array}
\usepackage{makecell}
\usepackage{diagbox}
\usetikzlibrary{arrows}

\usepackage{tabularx}

\usepackage{caption}
\usepackage{subcaption}
\usepackage{dcolumn}
\newcolumntype{d}[1]{D{.}{.}{#1}}
\usepackage{multirow}
\usepackage{appendix}

\usepackage[firstpage]{draftwatermark}
\SetWatermarkText{\hspace*{6in}\raisebox{7.2in}{\includegraphics{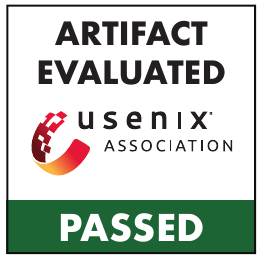}}}
\SetWatermarkAngle{0}

\usepackage{amsmath, amsfonts, amssymb, amsthm}
\usepackage{txfonts} 
\usepackage{verbatim}
\usepackage{url}
\usepackage{color}
\usepackage{algorithm2e}
\usepackage[noend]{algpseudocode}
\usepackage{multirow}

\usepackage{siunitx}
\ifx\BlackBox\undefined
\newcommand{\BlackBox}{\rule{1.5ex}{1.5ex}}  
\fi
\ifx\proof\undefined
\newenvironment{proof}{\par\noindent{\bf Proof\ }}{\hfill\BlackBox\\[2mm]}
\fi

\newcommand{\shortgap}{\vspace{6pt}}
\newcommand\shortsection[1]{\vspace{6pt}{\noindent\bf #1.}}
\newcommand\shortersection[1]{\vspace{6pt}{\noindent\em #1.}}

\theoremstyle{definition}

\makeatletter
\def\url@leostyle{%
  \@ifundefined{selectfont}{\def\UrlFont{\sf}}{\def\UrlFont{\small\sffamily}}}
\makeatother
\makeatletter
\def\url@beostyle{%
  \@ifundefined{selectfont}{\def\UrlFont{\sf}}{\def\UrlFont{\scriptsize\sffamily}}}
\makeatother

\usepackage{cleveref}
\LinesNumbered
\newcommand\candidate[1]{\ensuremath{{#1}^{\prime}}}
\newcommand\adversarial[1]{\ensuremath{{#1}^{\star}}}

\usepackage{setspace}
\DeclareMathOperator*{\argmin}{arg\,min}
\usepackage{listings}
\usepackage{multicol}
\newcommand\model[1]{{\sf\small #1}}
\newcommand{\NA}{\model{NA}}
\newcommand{\NB}{\model{NB}}
\newcommand{\NC}{\model{NC}}

\newcommand{\BanditsTD}{$\text{Bandits}_{\text{TD}}$}

\hypersetup{
    colorlinks=false,
    linkcolor=black,
    filecolor=black,      
    urlcolor=black,
}

\newcommand{\squishlist}{
   \begin{list}{$\bullet$}
    { \setlength{\itemsep}{0pt}      \setlength{\parsep}{3pt}
      \setlength{\topsep}{3pt}       \setlength{\partopsep}{0pt}
      \setlength{\leftmargin}{1.5em} \setlength{\labelwidth}{1em}
      \setlength{\labelsep}{0.5em} } }

\newcommand{\squishend}{
    \end{list}  }
    
\title{
\vspace*{-0.5in}                    
{{\normalsize \rm In 29\textsuperscript{th} {\em USENIX Security Symposium}, August 2020 (Accepted: June 2019; This version:  \today) \hrule}}
 \vspace*{0.4in}
Hybrid Batch Attacks: Finding Black-box\\Adversarial Examples with Limited Queries}

\author{
{\rm Fnu Suya, Jianfeng Chi, David Evans, Yuan Tian}\\
University of Virginia
}
\date{}
\begin{document}

\maketitle

\begin{abstract}
We study adversarial examples in a black-box setting where the adversary only has API access to the target model and each query is expensive. Prior work on black-box adversarial examples follows one of two main strategies: 
(1) transfer attacks use white-box attacks on local models to find candidate adversarial examples that transfer to the target model, and 
(2) optimization-based attacks use queries to the target model
and apply optimization techniques to search for adversarial examples. We propose hybrid attacks that combine both strategies, using candidate adversarial examples from local models as starting points for optimization-based attacks and using labels learned in optimization-based attacks to tune local models for finding transfer candidates. 
We empirically demonstrate on the MNIST, CIFAR10, and ImageNet datasets that our hybrid attack strategy reduces cost and improves success rates. We also introduce a seed prioritization strategy which enables attackers to focus their resources on the most promising seeds. Combining hybrid attacks with our seed prioritization strategy enables batch attacks that can reliably find adversarial examples with only a handful of queries. 
\end{abstract}

\section{Introduction}\label{Introduction}
Machine learning (ML) models are often prone to misclassifying inputs, known as \emph{adversarial examples} (AEs), that are crafted by perturbing a normal input in a constrained, but purposeful way. Effective methods for finding adversarial examples have been found in white-box settings, where an adversary has full access to the target model~\cite{szegedy2013intriguing, goodfellow2014explaining, carlini2017towards,madry2017towards,kurakin2016adversarial}, as well as in black-box settings, where only API access is available~\cite{chen2017zoo,chen2018AutoZOOM,papernot2017practical,suya2017query,ilyas2017query,ilyas2018prior}. 
In this work, we aim to improve our understanding of the expected cost of black-box attacks in realistic settings. For most scenarios where the target model is only available through an API, the cost of attacks can be quantified by the number of model queries needed to find a desired number of adversarial examples. Black-box attacks often require a large number of model queries, and each query takes time to execute, in addition to incurring a service charge and exposure risk to the attacker. 

Previous black-box attacks can be grouped into two categories: \emph{transfer attacks}~\cite{papernot2016transferability,papernot2017practical} and \emph{optimization attacks} ~\cite{suya2017query,chen2017zoo,chen2018AutoZOOM,ilyas2017query,ilyas2018prior}. Transfer attacks exploit the observation that adversarial examples often transfer between different models~\cite{liu2016delving,tramer2017ensemble,goodfellow2014explaining,papernot2017practical,li2018query}. The attacker generates adversarial examples against local models using white-box attacks, and hopes they transfer to the target model. Transfer attacks use one query to the target model for each attempted candidate transfer, but suffer from \emph{transfer loss} as local adversarial examples may not successfully transfer to the target model. Transfer loss can be very high, especially for targeted attacks where the attacker's goal requires finding examples where the model outputs a particular target class rather than just producing misclassifications.


Optimization attacks formulate the attack goal as a black-box optimization problem and carry out the attack using a series of queries to the target model~\cite{chen2017zoo,bhagoji2017exploring,chen2018AutoZOOM,ilyas2017query,ilyas2018prior,al2019there,guo2019simple,moon2019parsimonious,li2019nattack}. These attacks require many queries, but do not suffer from transfer loss as each seed is attacked interactively using the target model. Optimization-based attacks can have high attack success rates, even for targeted attacks, but often require many queries for each adversarial example found.

\shortsection{Contributions} 
Although improving query efficiency and attack success rates for black-box attacks is an active area of research for both transfer-based and optimization-based attacks, prior works treat the two types of attacks independently and fail to explore possible connections between the two approaches. 
We investigate three straightforward possibilities for combining transfer and optimization-based attacks (Section~\ref{sec:hybrid_attack}), and find that only one is generally useful (Section \ref{sec:hybrid_evaluation}): failed transfer candidates are useful starting points for optimization attacks. This can be used to substantially improve black-box attacks in terms of both success rates and, most importantly, query cost.
Compared to transfer attacks, hybrid attacks can significantly improve the attack success rate by adopting optimization attacks for the non-transfers, which increases per-sample query cost. Compared to optimization attacks, hybrid attacks significantly reduce query complexity when useful local models are available. For example, for both MNIST and CIFAR10, our hybrid attacks reduce the mean query cost of attacking normally-trained models by over 75\% compared to state-of-the-art optimization attacks. For ImageNet, the transfer attack only has 3.4\% success rate while the hybrid attack approaches 100\% success rate.

To improve our understanding of resource-limited black-box attacks, we simulate a \emph{batch attack} scenario where the attacker has access to a large pool of seeds and is motivated to obtain many adversarial examples using limited resources. Alternatively, we can view the batch attacker's goal as obtaining a fixed number of adversarial examples with fewest queries. We demonstrate that the hybrid attack can be combined with a novel seed prioritization strategy to dramatically reduce the number of queries required in batch attacks (Section~\ref{sec:batch_attack}). For example, for ImageNet, when the attacker is interested in obtaining 10 adversarial examples from a pool of 100 candidate seeds, our seed prioritization strategy can be used to save over 70\% of the queries compared to random ordering of the seeds.    

\begin{table*}[h]
\centering
\def\arraystretch{1.8}
\scalebox{1}{\begin{tabular}{cccc}
\toprule
Attack & Gradient Estimation & Queries per Iteration & White-box Attack \\ \hline
ZOO~\cite{chen2017zoo} &  $\mathbf{\hat{g}} = \{\hat{g}_{i},\hat{g}_{2},...,\hat{g}_{D}\}$,
$\hat{g}_{i}\approx \frac{f(\mathbf{x}+\delta e_{i})-f(\mathbf{x}-\delta e_{i})}{\delta}$ & $2D$ & CW~\cite{carlini2017towards} \\ \hline
Bhagoji et. al~\cite{bhagoji2017exploring}&  
ZOO + random feature group or PCA & $\leq 2D$ & FGSM~\cite{goodfellow2014explaining}, PGD~\cite{madry2017towards} \\ \hline
AutoZOOM~\cite{chen2018AutoZOOM} & 
\multicolumn{1}{c}{$\mathbf{u}_{i}\sim U$, $\mathbf{\hat{g}} = \frac{1}{N}\sum_{i}^{N}\frac{f(\mathbf{x}+\delta \mathbf{u}_{i})-f(\mathbf{x})}{\delta}\mathbf{u}_{i} $} & $N+1$ & CW~\cite{carlini2017towards} \\ \hline
NES~\cite{ilyas2017query} 
& $\mathbf{u}_{i}\sim \mathcal{N}(\mathbf{0},\mathbf{\it{I}})$, $\mathbf{\hat{g}} = \frac{1}{N}\sum_{i}^{N} \frac{f(\mathbf{x}+\delta \mathbf{u}_{i})}{\delta}\mathbf{u}_{i} $ & $N$ & PGD \\ \hline
\BanditsTD\ \cite{ilyas2018prior}& NES + time/data dependent info & $N$ & PGD \\ \hline
SignHunter~\cite{al2019there}&  Gradient sign w/ divide-and-conquer method & $2^{\left \lceil{\log(D)+1}\right \rceil}$ & PGD \\ \hline
Cheng et al.~\cite{cheng2019improving}&  
$\mathbf{u}_{i}\sim U,\mathbf{\hat{g}} = \frac{1}{N}\sum_{i}^{N} (\sqrt{\lambda}\cdot \mathbf{v} + \sqrt{1-\lambda}\cdot \frac{(\mathbf{I}-\mathbf{v}\mathbf{v}^{T})\mathbf{u}_i}{||(\mathbf{I}-\mathbf{v}\mathbf{v}^{T})\mathbf{u}_{i}||_{2}})$ 
 & $N$ & PGD \\ \hline
\end{tabular}
}
\caption{Gradient attacks. These attacks use some method to estimate gradients and then leverage white-box attacks. $D$ is data dimension, $e_{i}$ denotes standard basis, $N$ is the number of gradient averages. $f(\mathbf{x})$ denotes prediction confidence of image $\mathbf{x}$: for targeted attacks, it denotes the confidence of target class; for untargeted attacks, it denotes the confidence of original class. $\delta$ is a small constant. $\mathbf{v}$ is the local model gradient. $\lambda$ is a constant controlling the strength of local and target model gradients.}
\label{tab:grad-methods}
\end{table*}
\section{Background and Related Work}
In this section, we overview the two main types of black-box attacks which are combined in our hybrid attack strategies.

\subsection{Transfer Attacks}\label{sec:background:transferattacks}
Transfer attacks take advantage of the observation that adversarial examples often transfer across models. The attacker runs standard white-box attacks on local models to find adversarial examples that are expected to transfer to the target model. Most works assume the attacker has access to similar training data to the data used for the target model, or has access to pretrained models for similar data distribution. For attackers with access to pretrained local models, no queries are needed to the target model to train the local models. 
Other works consider training a local model by querying the target model, sometimes referred to as \emph{substitute training}~\cite{papernot2017practical,li2018query}. With na\"{i}ve substitute training, many queries are needed to train a useful local model. Papernot et al.\ adopt a reservoir sampling approach to reduce the number of queries needed~\cite{papernot2017practical}. Li et al.\ use active learning to further reduce the query cost~\cite{li2018query}. However, even with these improvements, many queries are still needed and substitute training has had limited effectiveness for complex target models.

Although adversarial examples sometimes transfer between models, transfer attacks typically have much lower success rates than optimization attacks, especially for targeted attacks. In our experiments on ImageNet, the highest transfer rate of targeted attacks observed from a single local model is 0.2\%, while gradient-based attacks achieve nearly 100\% success. Liu et al.\ improve transfer rates by using an ensemble of local models~\cite{liu2016delving}, but still only achieve low transfer rates (3.4\% in our ImageNet experiments, see Table \ref{tab:hypo_loc_adv_sample}). 

Another line of work aims to improve transferability by modifying the white-box attacks on the local models. Dong et al.\ adopt the momentum method to boost the attack process and leads to improved transferability \cite{dong2018boosting}. Xie et al.\ improve the diversity of attack inputs by considering image transformations in the attack process to improve transferability of existing white-box attacks~\cite{xie2018improving}. Dong et al.\ recently proposed a translation invariant optimization method that further improves transferability~\cite{dong2019evading}. We did not incorporate these methods in our experiments, but expect they would be compatible with our hybrid attacks.

\begin{table*}[th]
\centering
\def\arraystretch{1.8}
\scalebox{1.0}{
\begin{tabular}{cccl}
\toprule
Attack & Applicable Norm & Objective Function & \multicolumn{1}{c}{Solution Method} \\ \hline
Sim-BA~\cite{guo2019simple} & $L_{2}, L_{\infty}$ & $\min \limits_{\mathbf{x}'}~f(\mathbf{x}')$ & Iterate: sample $\mathbf{q}$ from $Q$, first try $\epsilon\mathbf{q}$, then $-\epsilon \mathbf{q}$ \\ \hline
$\mathcal{N}$Attack ~\cite{li2019nattack} & $L_{2}, L_{\infty}$ & \begin{tabular}[c]{@{}c@{}} $\min \limits_{\theta}\int l(\mathbf{x}')\pi(\mathbf{x}'|\theta)d\mathbf{x}' $ \end{tabular}& Compute $\theta^{*}$, then sample from $\pi(\mathbf{x}'\;|\;\theta^{*})$ \\ \hline
Moon et al.~\cite{chen2018AutoZOOM} &$L_{\infty}$ &  $\max\limits_{\mathcal{S}\subseteq \mathcal{V}}f(\mathbf{x}+\epsilon\sum_{i\in\mathcal{S}}e_{i}-\epsilon\sum_{i\notin \mathcal{S}}e_{i})$ & Compute $\mathcal{S}^{*}$, then $\mathbf{x}+\epsilon\sum_{i\in \mathcal{S}^{*}}e_{i}-\epsilon\sum_{i\notin \mathcal{S}^{*}}e_{i}$ \\ \hline
\end{tabular}
}
\caption{Gradient-free attacks. These attacks define an objective function and obtain the AE by solving the optimization problem. $Q$ denotes a set of orthonormal candidate vectors, $l(\mathbf{x}')$ denotes the cross-entropy loss of image $\mathbf{x}^{'}$ with original label (untargeted attack) or target label (targeted attack). $\pi(\mathbf{x}'|\theta)$ denotes the distribution of $\mathbf{x}'$ parameterized by $\theta$, $\mathcal{V}$ denotes ground set of all pixel locations. Variables with $*$ are locally-optimal solutions obtained by solving the corresponding optimization problems.}
\label{tab:non-grad-methods}
\end{table*}


\subsection{Optimization Attacks}\label{sec:background:bbattacks}
Optimization-based attacks work by defining an objective function and iteratively perturbing the input to optimize that objective function. We first consider optimization attacks where the query response includes full prediction scores, and categorize those ones that involve estimating the gradient of the objective function using queries to the target model, and those that do not depend on estimating gradients. Finally, we also briefly review restricted black-box attacks, where attackers obtain even less information from each model query, in the extreme, learning just the label prediction for the test input.


\shortsection{Gradient Attacks}
Gradient-based black-box attacks numerically estimate the gradient of the target model, and execute standard white-box attacks using those estimated gradients. Table~\ref{tab:grad-methods} compares several gradient black-box attacks.

The first attack of this type was the ZOO (zeroth-order optimization) attack, introduced by 
Chen et al.\ \cite{chen2017zoo}. It adopts the finite-difference method with dimension-wise estimation to approximate gradient values, and uses them to execute a Carlini-Wagner (CW) white-box attack~\cite{carlini2017towards}. The attack runs for hundreds to thousands of iterations and takes $2D$ queries per CW optimization iteration, where $D$ is the dimensionality. Hence, the query cost is extremely high for larger images (e.g., over 2M queries on average for ImageNet). 

Following this work, several researchers have sought more query-efficient methods for estimating gradients for executing black-box gradient attacks. Bhagoji et al.\ propose reducing query cost of dimension-wise estimation by randomly grouping features or estimating gradients along with the principal components given by principal component analysis (PCA)~\cite{bhagoji2017exploring}. Tu et al.'s AutoZOOM attack uses two-point estimation based on random vectors and reduces the query complexity per CW iteration from $2D$ to $2$ without losing much accuracy on estimated gradients~\cite{chen2018AutoZOOM}. Ilyas et al.'s NES attack~\cite{ilyas2017query} uses a natural evolution strategy (which is in essence still random vector-based gradient estimation)~\cite{wierstra2008natural}, to estimate the gradients for use in projected gradient descent (PGD) attacks~\cite{madry2017towards}. 

Ilyas et al.'s \BanditsTD\ attack incorporates time and data dependent information into the NES attack~\cite{ilyas2018prior}. Al-Dujaili et al.'s SignHunter adopts a divide-and-conquer approach to estimate the sign of the gradient and is empirically shown to be superior to the \BanditsTD\ attack in terms of query efficiency and attack success rate~\cite{al2019there}. Cheng et al.\ recently proposed improving the \BanditsTD\ attack by incorporating gradients from surrogate models as priors when estimating the gradients~\cite{cheng2019improving}. For our experiments (Section \ref{subsec:setupattacks}), we use AutoZOOM and NES as representative state-of-the-art black-box attacks.\footnote{We also tested \BanditsTD\ on ImageNet, but found it less competitive to the earlier attacks and therefore, do not include the results in this paper. We have not evaluated SignHunter and the attack of Cheng et al.~\cite{cheng2019improving}, but plan to include more results in the future versions and have released an open-source framework to enable other attacks to be tested using our methods.} 

\shortsection{Gradient-free Attacks}
\label{sec:non_grad_attack}
Researchers have also explored search-based black-box attacks using {\em heuristic} methods that are not based on gradients, which we call gradient-free attacks.  One line of work directly applies known heuristic black-box optimization techniques, and is not competitive with the gradient-based black-box attacks in terms of query efficiency. 
Alzantot et al.~\cite{alzantot2018genattack} develop a genetic programming strategy, where the fitness function is defined similarly to CW loss~\cite{carlini2017towards}, using the prediction scores from queries to the black-box model. A similar genetic programming strategy was used to perform targeted black-box attacks on audio systems~\cite{taori2018targeted}. 
Narodytska et al.\ \cite{narodytska2016simple} use a local neighbor search strategy, where each iteration perturbs the most significant pixel. Since the reported query efficiency of these methods is not competitive with results for gradient-based attacks, we did not consider these attacks in our experiments.

Several recent gradient-free black-box attacks (summarized in Table~\ref{tab:non-grad-methods}) have been proposed that can significantly outperform the gradient-based attacks. Guo et al.'s Sim-BA~\cite{guo2019simple} iteratively adds or subtracts a random vector sampled from a predefined set of orthonormal candidate vectors to generate adversarial examples efficiently. Li et al.'s 
$\mathcal{N}$Attack~\cite{li2019nattack} formulates the adversarial example search process as identifying a probability distribution from which random samples are likely to be adversarial. 
Moon et al.\ formulate the $L_{\infty}$-norm black-box attack with $\epsilon$ perturbation as a problem of selecting a set of pixels with $+\epsilon$ perturbation and applying the $-\epsilon$ perturbation to the remaining pixels, such that the objective function defined for misclassification becomes a set maximization problem. Efficient submodular optimization algorithms are then used to solve the set maximization problem efficiently~\cite{moon2019parsimonious}.  These attacks became available after we started our experiments, so are not included in our experiments. However, our hybrid attack strategy is likely to work for these new attacks as it boosts the optimization attacks by providing better starting points, which we expect is beneficial for most attack algorithms. 

\shortsection{Restricted Black-box Attacks}\label{sec:other_grad_attacks}
All the previous attacks assume the adversary can obtain complete prediction scores from the black-box model. Much less information might be revealed at each model query, however, such as just the top few confidence scores or, at worst, just the output label.

Ilyas et al.~\cite{ilyas2017query}, in addition to their main results of NES attack with full prediction scores, also consider scenarios where prediction scores of the top-$k$ classes or only the model prediction label are revealed. In the case of partial prediction scores, attackers start from an instance in the target class (or class other than the original class) and gradually move towards the original image with the estimated gradient from NES. For the label-only setting, a surrogate loss function is defined to utilize the strategy of partial prediction scores. Brendel et al.~\cite{brendel2017decision} propose a label-only black-box attack, which starts from an example in the target class and performs a random walk from that target example to the seed example. This random walk procedure often requires many queries. Following this work, several researchers have worked to reduce the high query cost of random walk strategies. Cheng et al.\ formulate a label-only attack as an optimization problem, reducing the query cost significantly compared to the random walk~\cite{cheng2018query}. Chen et al.\ also formulate the label-only attack as an optimization problem and show this significantly improves query efficiency~\cite{chen2019boundary}. Brunner et al.~\cite{brunner2018guessing} improve upon the random walk strategy by additionally considering domain knowledge of image frequency, region masks and gradients from surrogate models.

In our experiments, we assume attackers have access to full prediction scores, but we believe our methods are also likely to help in settings where attackers obtain less information from each query. This is because the hybrid attack boosts gradient attacks by providing better starting points and is independent from the specific attack methods or the types of query feedback from the black-box model.

\section{Hybrid Attacks}\label{sec:hybrid_attack}

Our hybrid attacks combine the transfer and optimization methods for searching for adversarial examples. Here, we introduce the threat model of our attack, state the hypotheses underlying the attacks, and presents the general hybrid attack algorithm. We evaluate the hypotheses and attacks in Section~\ref{sec:hybrid_evaluation}.

\shortsection{Threat Model} In the black-box attack setting, the adversary does not have direct access to the target model or knowledge of its parameters, but can use API access to the target model to obtain prediction confidence scores for a limited number of submitted queries. We assume the adversary has access to pretrained local models for the same task as the target model. These could be directly available or produced from access to similar training data and knowledge of the model architecture of the target model. The assumption of having access to pretrained local models is a common assumption for research on transfer-based attacks. A few works on substitute training~\cite{papernot2017practical, li2018query} have used weaker assumptions such as only having access to a small amount of training data, but have only been effective so far for very small datasets. 

\shortsection{Hypotheses}
Our approach stems from three hypotheses about the nature of adversarial examples: 

\shortersection{Hypothesis 1 (H1): Local adversarial examples are better starting points for optimization attacks than original seeds}
Liu et al.\ observe that for the same classification tasks, different models tend to have similar decision boundaries \cite{liu2016delving}. Therefore, we hypothesize that, although candidate adversarial examples generated on local models may not fully transfer to the target model, these candidates are still closer to the targeted region than the original seed and hence, make better starting points for optimization attacks. 

\shortersection{Hypothesis 2 (H2): Labels learned from optimization attacks can be used to tune local models}
Papernot et al.\ observe that generating examples crossing decision boundaries of local models can produce useful examples for training local models closer to the target model~\cite{papernot2017practical}. Therefore, we hypothesize that query results generated through the optimization search queries may contain richer information regarding true target decision boundaries. These new labeled inputs that are the by-product of an optimization attack can then be used to fine-tune the local models to improve their transferability. 

\shortersection{Hypothesis 3 (H3): Local models can help direct gradient search} Since different models tend to have similar decision boundaries for the same classification tasks, we hypothesize that gradient information obtained from local models may also help better calibrate the estimated gradient of gradient based black-box attacks on target model. 

\shortgap

We are not able to find any evidence to support the third hypothesis (H3), which is consistent with Liu et al.'s results~\cite{liu2016delving}. They observed that, for ImageNet models, the gradients of local and target models are almost orthogonal to each other. 
We also tested this for MNIST and CIFAR10, conducting white-box attacks on local models and storing the intermediate images and the corresponding gradients. We found that the local and target models have almost orthogonal gradients (cosine similarity close to zero) and therefore, a na\"{i}ve combination of gradients of local and target model is not feasible. One possible explanation is the noisy nature of gradients of deep learning models, which causes the gradient to be highly sensitive to small variations~\cite{balduzzi2017shattered}. {Although the cosine similarity is low, two recent works have attempted to combine the local gradients and the estimated gradient of the black-box model by a linear combination~\cite{cheng2019improving,brunner2018guessing}. However, Brunner et al.\ observe that straightforward incorporation of local gradients does not improve targeted attack efficiency much~\cite{brunner2018guessing}. Cheng et al.\ successfully incorporated local gradients into untargeted black-box attacks, however, they
do not consider the more challenging targeted attack scenario and it is still unclear if local gradients can help in more challenging cases~\cite{brunner2018guessing}.} Hence, we do not investigate this further in this paper and leave it as an open question if there are more sophisticated ways to exploit local model gradients.

\shortsection{Attack Method} \label{subsec:hybrid_method}
Our hybrid attacks combine transfer and optimization attacks in two ways based on the first two hypotheses: we use a local ensemble to select better starting points for an optimization attack, and use the labeled inputs obtained in the optimization attack to tune the local models to improve transferability. Algorithm~\ref{alg:hybrid} provides a general description of the attack. The attack begins with a set of seed images $\mathbf{X}$, which are natural images that are correctly classified by the target model, and a set of local models, $F$. The attacker's goal is to find a set of successful adversarial examples (satisfying some attacker goal, such as being classified in a target class with a limited perturbation below starting from a natural image in the source class).

\begin{algorithm}[bt]
\SetKwInOut{Input}{input}\SetKwInOut{Output}{output}
    \Input{Set of seed images $\mathbf{X}$ with labels, \\ 
    local model ensemble $F$, \\ target black-box model $g$}
    \Output{Set of successful adversarial examples}
    
    $\mathbf{R} \leftarrow \mathbf{X}$ {\em (remaining seeds to attack)} \\
    $A \leftarrow \emptyset$ {\em(successful adversarial examples)} \\
    $\mathbf{Q} \leftarrow \mathbf{X}$ {\em (fine-tuning set for local models)} \label{line:tuneset}\\
    \While{$\mathbf{R}$ is not empty}{
        select and remove the next seed to attack \\
        $\mathbf{x} \leftarrow \text{\em selectSeed}(\mathbf{R}, F)$ \label{line:selectseed}\\
        use local models to find a candidate adversarial example\\
        $\candidate{\mathbf{x}} \leftarrow \text{\em whiteBoxAttack}(F, \mathbf{x})$ \\
        {
        $\adversarial{\mathbf{x}},\; S \leftarrow \text{\em blackBoxAttack}(\mathbf{x}, \candidate{\mathbf{x}},g)$ \label{line:bbattack} 
        \\
        \If{\adversarial{\mathbf{x}}}{
           $A.insert(<\mathbf{x},\; \adversarial{\mathbf{x}}>)$
        }
        $\mathbf{Q}.insert($S$)$ \label{line:auugmenttuneset}\\
        use byproduct labels to retrain local models
\\      $tuneModels(F,\mathbf{Q})$ \label{line:tunemodels} \\
}
    } 
    \Return $A$ \\
\caption{Hybrid Attack.} \label{alg:hybrid}
\end{algorithm}

The attack proceeds by selecting the next seed to attack (line~\ref{line:selectseed}). Section~\ref{sec:hybrid_evaluation} considers the case where the attacker only selects seeds randomly; Section~\ref{sec:batch_attack} considers ways more sophisticated resource-constrained attackers may improve efficiency by prioritizing seeds. Next, the attack uses the local models to find a candidate adversarial example for that seed. When the local adversarial example is found, we first check its transferability and if the seed directly transfers, we proceed to attack the next seed. If the seed fails to directly transfer, the black-box optimization attack is then executed starting from that candidate. The original seed is also passed into the black-box attack (line~\ref{line:bbattack}) since the adversarial search space is defined in terms of the original seed $\mathbf{x}$, not the starting point found using the local models, \candidate{\mathbf{x}}.  This is because the space of permissible inputs is defined based on distance from the original seed, which is a natural image. Constraining with respect to the space of original seed is important because we need to make sure the perturbations from our method are still visually indistinguishable from the natural image. If the black-box attack succeeds, it returns a successful adversarial example, \adversarial{\mathbf{x}}, which is added to the returned set. Regardless of success, the black-box attack produces input-label pairs ($S$) during the search process which can be used to tune the local models (line~\ref{line:tunemodels}), as described in Section~\ref{sec:loc_model_tune}.

\section{Experimental Evaluation}\label{sec:hybrid_evaluation}

In this section, we report on experiments to validate our hypothesis, and evaluate the hybrid attack methods. 
Section~\ref{subsec:setupmodels} describes the experimental setup; Section~\ref{subsec:setupattacks} describes the attack configuration; Section~\ref{subsec:attackgoal} describes the attack goal; Section~\ref{subsec:valid_hypo} reports on experiments to test the first hypothesis from Section~\ref{sec:hybrid_attack} and measure the effectiveness of hybrid attacks; Section~\ref{sec:improve_robust_attack} improves the attack for targeting robust models, and Section~\ref{sec:loc_model_tune} evaluates the second hypothesis, showing the impact of tuning the local models using the label byproducts.
For all of these, we focus on comparing the cost of the attack measured as the average number of queries needed per adversarial example found across a set of seeds. In Section~\ref{sec:batch_attack}, we revisit the overall attack costs in light of batch attacks that can prioritize which seeds to attack. 

\subsection{Datasets and Models}\label{subsec:setupmodels}

We evaluate our attacks on three popular image classification datasets and a variety of state-of-the-art models. 

\shortsection{MNIST} MNIST~\cite{lecun1998mnist} is a dataset of 70,000 $28\times28$ greyscale images of handwritten digits (0--9), split into 60,000 training and 10,000 testing samples. 
For our normal (not adversarially trained) MNIST models, we use the pretrained MNIST models of Bhagoji et al.~\cite{bhagoji2017exploring}, which typically consist of convolutional layers and fully connected layers. We use their MNIST model A as the target model, and models B--D as local ensemble models. To consider the more challenging scenario of attacking a black-box robust model, we use Madry's robust MNIST model, which demonstrates strong robustness even against the best white-box attacks (maintaining over 88\% accuracy for $L_{\infty}$ attacks with $\epsilon=0.3$)~\cite{madry2017towards}.

\shortsection{CIFAR10} CIFAR10 \cite{krizhevsky2009learning} consists of 60,000 $32\times32$ RGB images, with 50,000 training and 10,000 testing samples for object classification (10 classes in total). We train a standard DenseNet model and obtain a test accuracy of 93.1\%, which is close to state-of-the-art performance. To test the effectiveness of our attack on robust models, we use Madry's CIFAR10 Robust Model \cite{madry2017towards}. Similarly, we also use the normal CIFAR10 target model and the standard DenseNet (Std-DenseNet) model interchangeably. For our normal local models, we adopt three simple LeNet structures~\cite{lecun1998gradient}, varying the number of hidden layers and hidden units.\footnote{We also tested with deep CNN models as our local ensembles. However, they provide only slightly better performance compared to simple CIFAR10 models, while the fine-tuning cost is much higher.} For simplicity, we name the three normal models \NA, \NB\ and \NC\ where \NA\ has the fewest parameters and \NC\ has the most parameters. To deal with the lower effectiveness of attacks on robust CIFAR10 model (Section \ref{subsec:valid_hypo}), we also adversarially train two deep CIFAR10 models (DenseNet, ResNet) similar to the Madry robust model as robust local models. The adversarially-trained DenseNet and ResNet models are named \model{R-DenseNet} and \model{R-ResNet}. 

\shortsection{ImageNet}
ImageNet~\cite{imagenet_cvpr09} is a dataset closer to real-world images with 1000 categories, commonly used for evaluating state-of-the-art deep learning models. We adopt the following pretrained ImageNet models for our experiments: ResNet-50~\cite{he2016deep}, DenseNet~\cite{huang2017densely}, VGG-16, and VGG-19~\cite{simonyan2014very} (all from \url{https://keras.io/applications/}). We take DenseNet as the target black-box model and the remaining models as the local ensemble.

\subsection{Attack Configuration}\label{subsec:setupattacks}

For the hybrid attack, since we have both the target model and local model, we have two main design choices: (1) which white-box attacks to use for the local models , and (2) which optimization attacks to use for the target model.

\shortsection{Local Model Configurations} \label{subsec:loc_model_config}
We choose an ensemble of local models in our hybrid attacks. This design choice is motivated by two facts: First, different models tend to have significantly different direct transfer rates to the same target model (see Figure~\ref{fig:local_ensembles}), when evaluated individually. Therefore, taking an ensemble of several models helps avoid ending up with a single local model with a very low direct transfer rate. Second, consistent with the findings of Liu et al.~\cite{liu2016delving} on attacking an ensemble of local models, for MNIST and CIFAR10, we find that the ensemble of normal local models yields the highest transfer rates when the target model is a normally trained model (note that this does not hold for robust target model, as shown in Figure~\ref{fig:local_ensembles} and discussed further in Section~\ref{sec:improve_robust_attack}). We validate the importance of normal local ensemble against normal target model by considering different combinations of local models (i.e., ${N \choose k}, k = 1,...,N$) and checking their corresponding transfer rates and the average query cost. We adopt the same approach as proposed by Liu et al.~\cite{liu2016delving} to attack multiple models simultaneously, where the attack loss is defined as the sum of the individual model loss. In terms of transfer rate, we observe that a single CIFAR10 or MNIST normal model can achieve up to 53\% and 35\% targeted transfer rate respectively, while an ensemble of local models can achieve over 63\% and 60\% transfer rate. In terms of the average query cost against normal target models, compared to a single model, an ensemble of local models on MNIST and CIFAR10 can save on average 53\% and 45\% of queries, respectively. Since the ensemble of normal local models provides the highest transfer rate against normal target models, to be consistent, we use that configuration in all our experiments attacking normal models. We perform white-box PGD~\cite{madry2017towards} attacks (100 iterative steps) on the ensemble loss. We choose the PGD attack as it gives a high transfer rate compared to the fast gradient sign method (FGSM) method~\cite{goodfellow2014explaining}. 

\shortsection{Optimization Attacks} We use two state-of-the-art gradient estimation based attacks in our experiments: NES, a natural evolution strategy based attack~\cite{ilyas2017query} and AutoZOOM, an autoencoder-based zeroth-order optimization attack~\cite{chen2018AutoZOOM} (see Section~\ref{sec:background:bbattacks}).
These two methods are selected as all of them are shown to improve upon \cite{chen2017zoo} significantly in terms of query efficiency and attack success rate. We also tested with the \BanditsTD\ attack, an improved version of the NES attack that additionally incorporates time and data dependent information~\cite{ilyas2018prior}. However, we find that \BanditsTD\ is not competitive with the other two attacks in our attack scenario and therefore we do not include its results here.\footnote{For example, for the targeted attack on ImageNet, the baseline \BanditsTD\ attack only has 88\% success rate and average query cost of 51,745, which are much worse than the NES and AutoZOOM attacks.} 
Both tested attacks follow an attack method that attempts queries for a given seed until either a successful adversarial example is found or the set maximum query limit is reached, in which case they terminate with a failure. For MNIST and CIFAR10, we set the query limit to be 4000 queries for each seed. AutoZOOM sets the default maximum query limit for each as 2000, however as we consider a harder attacker scenario (selecting least likely class as the target class), we decide to double the maximum query limit. NES does not contain evaluation setups for MNIST and CIFAR10 and therefore, we choose to enforce the same maximum query limit as AutoZOOM.\footnote{By running the original AutoZOOM attack with a 4000 query limit compared to their default setting of 2000, we found 17.2\% and 25.4\% more adversarial samples out of 1000 seeds for CIFAR10 and MNIST respectively.} For ImageNet, we set the maximum query limit as 10,000 following the default setting used in the NES paper~\cite{ilyas2017query}. 

\begin{table*}[tb]
\centering
\begingroup
\setlength{\tabcolsep}{5pt} 
\scalebox{1}{
\begin{tabular}{ccd{4.0}crrrrrrrr}


\toprule
\multirow{2}{*}{Dataset} & 
Target & 
\multicolumn{1}{c}{Transfer} & 
Gradient & 
\multicolumn{2}{c}{Success (\%)} &
\multicolumn{2}{c}{Queries/Seed} & 
\multicolumn{2}{c}{Queries/AE} & 
\multicolumn{2}{c}{Queries/Search} 
\\ 
& Model & \multicolumn{1}{c}{Rate (\%)} & Attack & \multicolumn{1}{r}{\em Base} & \multicolumn{1}{r}{\em Ours} & \multicolumn{1}{r}{\em Base} & \multicolumn{1}{r}{\em Ours} & \multicolumn{1}{r}{\em Base} & \multicolumn{1}{r}{\em Ours} & \multicolumn{1}{r}{\em Base} & \multicolumn{1}{r}{\em Ours} \\ \hline

\multirow{4}{*}{MNIST} & \multirow{2}{*}{$ \text{Normal (T)}$} & \multirow{2}{*}{62.8} & AutoZOOM & 91.3 & \textbf{98.9} & 1,471 & \textbf{279} & 1,610 & \textbf{282} & 3,248 & \textbf{770} \\ 
 &  &  & NES & 77.5 & \textbf{89.2} & 2,544 & \textbf{892} & 3,284 & \textbf{1,000} & 8,254 & \textbf{3,376} \\ \cline{2-12} 
 & \multirow{2}{*}{$\text{Robust (U)}$} & \multirow{2}{*}{3.1} & 
 
 AutoZOOM & 7.5 & \textbf{7.5} & 3,755 & \textbf{3,748} & 50,102  & \textbf{49,776} & \textbf{83,042} & {83,806} \\ 
 &  &  & NES & 4.7 & \textbf{5.5} & 3,901 & \textbf{3,817} & {83,881} & \textbf{69,275} & 164,302 & \textbf{160,625} \\ \hline 
\multirow{4}{*}{CIFAR10} & \multirow{2}{*}{$\text{Normal (T)}$} & \multirow{2}{*}{63.6} & AutoZOOM & 92.9 & \textbf{98.2} & 1,117 & \textbf{271} & 1,203 & \textbf{276} & 2,143 & \textbf{781} \\ 
 &  &  & NES & 98.8 & \textbf{99.8} & 1,078 & \textbf{339} & 1,091 & \textbf{340} & 1,632 & \textbf{934} \\ \cline{2-12} 
 & \multirow{2}{*}{$\text{Robust (U)}$} & \multirow{2}{*}{10.1} & AutoZOOM & 64.3 & \textbf{65.3} & 1,692 & \textbf{1,652} & 2,632 & \textbf{2,532} & 3,117 & \textbf{2,997} \\ 
 &  &  & NES & \textbf{38.1} & {38.0} & 2,808 & \textbf{2,779} & 7,371 & \textbf{7,317} & \textbf{9,932} & {9,977} \\ \hline 
\multirow{2}{*}{ImageNet} & \multirow{2}{*}{$\text{Normal (T)}$} & \multirow{2}{*}{3.4} & AutoZOOM & 95.4 & \textbf{98.0} & 42,310 & \textbf{29,484} & 44,354 & \textbf{30,089} & 45,166 & \textbf{31,174} \\ 
 &  &  & NES & 100.0 & \textbf{100.0} & 18,797 & \textbf{14,430} & 18,797 & \textbf{14,430} & 19,030 & \textbf{14,939} \\ \bottomrule 
\end{tabular}
}
\caption{Impact of starting from local adversarial examples (Hypothesis 1). Baseline attacks that start from the original seeds are \emph{Base}; the hybrid attacks that start from local adversarial examples are \emph{Ours}. The attacks against the normal models are targeted (T), and against the robust models are untargeted (U). The \emph{Transfer Rate} is the direct transfer rate for local adversarial examples. The \emph{Success} rate is the fraction of seeds for which an adversarial example is found. The \emph{Queries/Seed} is the average number of queries per seed, regardless of success. The \emph{Queries/AE} is the average number of queries per successful adversarial example found, which is our primary metric. The \emph{Queries/Search} is the average number of queries per successful AE found using the gradient attack, excluding those found by direct transfer. Transfer attacks are independent from the subsequent gradient attacks and hence, transfer rates are separated from the specific gradient attacks. All results are averaged over 5 runs.
}
\label{tab:hypo_loc_adv_sample}
\endgroup
\end{table*}


\subsection{Attacker Goal}\label{subsec:attackgoal}
For MNIST and CIFAR10, we randomly select 100 images from each of the 10 classes for 1000 total images, against which we perform all black-box attacks. For ImageNet, we randomly sample 100 total images across all 1000 classes. 

\shortsection{Target Class}
We evaluate targeted attacks on the normal MNIST, CIFAR10, and ImageNet models. Targeted attacks are more challenging and are generally of more practical interest. For the MNIST and CIFAR10 datasets, all of the selected instances belong to one particular original class and we select as the target class the \emph{least likely class} of the original class given a prediction model, which should be the most challenging class to target. We define the least likely class of a class as the class which is most frequently the class with the lowest predicted probability across all instances of the class. For ImageNet, we choose the least likely class of each image as the target class. For the robust models for MNIST and CIFAR10, we evaluate untargeted attacks as these models are designed to resist untargeted attacks~\cite{madry_mnist_challenge,madry_cifar10_challenge}. Untargeted attacks against these models are significantly more difficult than targeted attacks against the normal models.

\shortsection{Attack Distance Metric and Magnitude}
We measure the perturbation distance using $L_{\infty}$, which is the most widely used attacker strength metric in black-box adversarial examples research. Since the AutoZOOM attack is designed for $L_2$ attacks, we transform it into an $L_{\infty}$ attack by clipping the attacked image into the $\epsilon$-ball ($L_{\infty}$ space) of the original seed in each optimization iteration. Note that the original AutoZOOM loss function is defined as $f(x)+ c\cdot \delta(x)$, where $f(x)$ is for misclassification (targeted or untargeted) and $\delta(x)$ is for perturbation magnitude minimization. In our transformation to $L_{\infty}$-norm, we only optimize $f(x)$ and clip the to $L_{\infty}$-ball of the original seed. NES is naturally an $L_{\infty}$ attack. For MNIST, we choose $\epsilon = 0.3$ following the setting in Bhagoji et al.~\cite{bhagoji2017exploring}. For CIFAR10, we set $\epsilon = 0.05$, following the same setting in early version of NES paper \cite{ilyas2017query}. For ImageNet, we set $\epsilon = 0.05$, as used by Ilyas et al.~\cite{ilyas2017query}.

\begin{table*}[th]
\scalebox{1.0}
\centering{
\begin{tabular}{ccccccccccc}
\toprule
\multirow{2}{*}{Target Model} & \multicolumn{2}{c}{Transfer Rate (\%)} & {Gradient} &  \multicolumn{2}{c}{Hybrid Success (\%)} & \multicolumn{2}{c}{Cost Reduction (\%)} & \multicolumn{2}{c}{Fraction Better (\%)} \\
 & \model{Normal-3} & \model{Robust-2} & Attack & \model{Normal-3} & \model{Robust-2} & \model{Normal-3} & \model{Robust-2} & \model{Normal-3} & \model{Robust-2} \\ \hline
\multirow{2}{*}{Normal} & \multirow{2}{*}{\textbf{63.6}} & \multirow{2}{*}{18.4} & AutoZOOM &\textbf{98.2} & 95.3 & \textbf{77.1} & 35.7 & \textbf{98.6} & 87.0 \\
 &  &  &  NES & \textbf{99.8} & 99.4 & \textbf{68.9} & 31.2 & \textbf{95.6} & 80.6 \\ \hline
\multirow{2}{*}{Robust} & \multirow{2}{*}{10.1} & \multirow{2}{*}{\textbf{40.7}} & AutoZOOM & 65.3 & \textbf{68.7} & 3.8 & \textbf{20.5} & 73.1 & \textbf{95.5} \\
 &  &  &  NES & 38.0 & \textbf{45.2} & 0.7 & \textbf{32.1} & 85.0 & \textbf{97.1} \\ 
 \bottomrule
\end{tabular}
}
\caption{Attack performance of all normal and all robust local ensembles on CIFAR10 target models. {The \model{Normal-3} ensemble is composed of the three normal models, \model{NA}, \model{NB}, and \model{NC}; the \model{Robust-2} ensemble is composed of \model{R-DenseNet} and \model{R-ResNet}. Results are averaged over 5 runs. Local model transfer rates are independent from the black-box attacks, so we separate transfer rate results from the black-box attack results.}}
\label{tab:cifar10_local_model_comp}\label{tab:cifar10_trans_local_model_comp}
\end{table*} 

\subsection{Local Candidates Results}\label{subsec:valid_hypo}

We test the hypothesis that local models produce useful candidates for black-box attacks by measuring the mean cost to find an adversarial example starting from both the original seed and from a candidate found using the local ensemble. All experiments are averaged over 5 runs to obtain more stable results. Table~\ref{tab:hypo_loc_adv_sample} summarizes our results. 

In nearly all cases, the cost is reduced by starting from the candidates instead of the original seeds, where candidates are generated by attacking local ensemble models. We measure the cost by the mean number of queries to the target model per adversarial example found. This is computed by dividing the total number of model queries used over the full attack on 1,000 (MNIST, CIFAR10) or 100 (ImageNet) seeds by the number of successful adversarial examples found. The overall cost is reduced by as much as 81\% (AutoZOOM attack on the normal MNIST model), and for both the AutoZOOM and for NES attack methods we see the cost drops by at least one third for all of the attacks on normal models (the improvements for robust models are not significant, which we return to in Section~\ref{sec:improve_robust_attack}). 
The cost drops for two reasons: some candidates transfer directly (which makes the query cost for that seed 1); others do not transfer directly but are useful starting points for the gradient attacks. To further distinguish the two factors, we include the mean query cost for adversarial examples found from non-transfering seeds as the last two columns in Table~\ref{tab:hypo_loc_adv_sample}. This reduction is significant for all the attacks across the normal models, up to 76\% (AutoZOOM attack on normal MNIST models).

The hybrid attack also offers success rates higher than the gradient attacks (and much higher success rates that transfer-only attacks), but with query cost reduced because of the directly transferable examples and boosting effect on gradient attacks from non-transferable examples. For the AutoZOOM and NES attacks on normally-trained MNIST models, the attack failure rates drop dramatically (from 8.7\% to 1.1\% for AutoZOOM, and from 22.5\% to 10.8\% for NES), as does the mean query cost (from 1,610 to 282 for AutoZOOM, and from 3,284 to 1,000 for NES). Even excluding the direct transfers, the saving in queries is significant (from 3,248 to 770 for AutoZOOM, and from 8,254 to 3,376 for NES). The candidate starting points are nearly always better than the original seed. For the two attacks on MNIST, there were only at most 28 seeds out of 1,000 where the original seed was a better starting point than the candidate; the worst result is for the AutoZOOM attack against the robust CIFAR10 model where 269 out of 1,000 of the local candidates are worse starting points than the corresponding original seed.

\begin{figure*}[tb]
        \centering
        \includegraphics[width=0.95\textwidth]{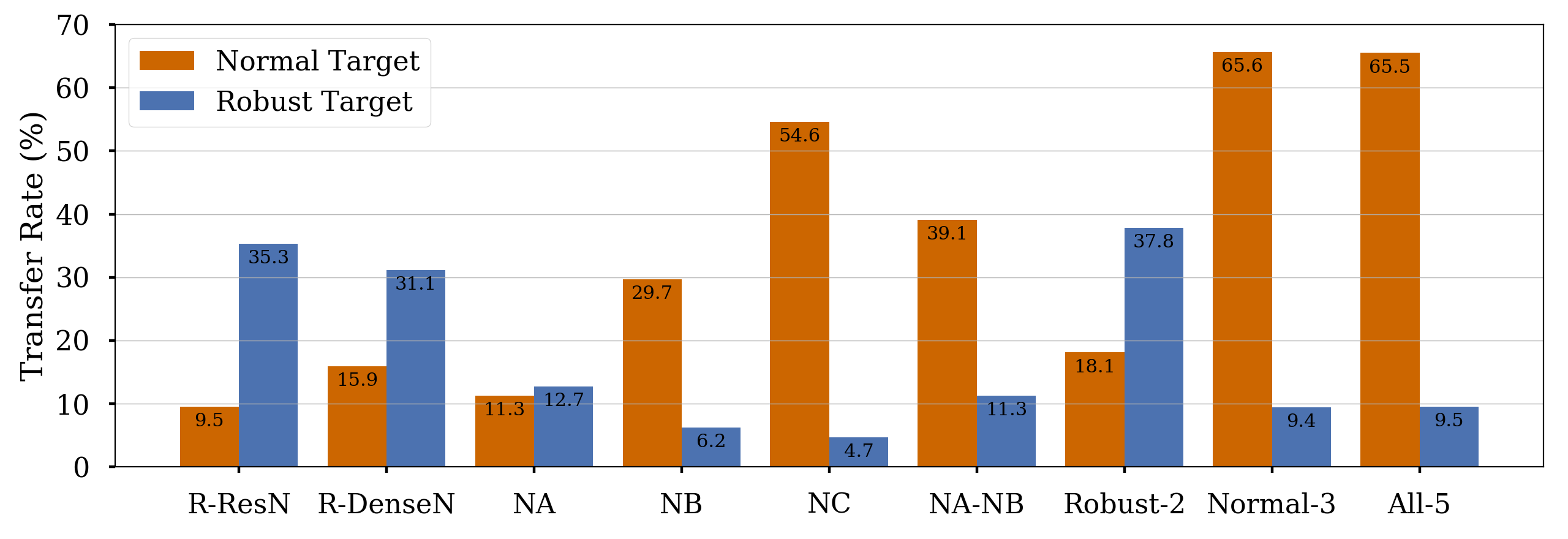}
        \caption[]%
        {Transfer rates of different local ensembles. {The \model{Normal-3} ensemble is composed of the three normal models, \model{NA}, \model{NB}, and \model{NC}; the \model{Robust-2} ensemble is composed of \model{R-DenseNet} and \model{R-ResNet}. 
        The \model{All-5} is composed of all of the 5 local models. Transfer rate is measured on independently sampled test images and is averaged over 5 runs.}}    
        \label{fig:local_ensembles}
\end{figure*} 
\subsection{Attacking Robust Models} \label{sec:improve_robust_attack}

The results in Table~\ref{tab:hypo_loc_adv_sample} show substantial improvements from hybrid attacks on normal models, but fail to provide improvements against the robust models. The improvements against robust models are less than 4\% for both attacks on both targets, except for NES against MNIST where there is $\sim$17\% improvement. We speculate that this is due to differences in the vulnerability space between normal and robust models, which means that the candidate adversarial examples found against the normal models in the local ensemble do not provide useful starting points for attacks against a robust model. This is consistent with Tsipras et al.'s finding that robust models for image classification tasks capture key features of images while normal models capture relatively noisy features~\cite{tsipras2018robustness}. Because of the differences in extracted features, adversarial examples against robust models require perturbing key features (of the target domain) while adversarial examples can be found against normal models by perturbing irrelevant features. This would explain why we did not see improvements from the hybrid attack when targeting robust models. To validate our hypothesis on the different attack surfaces, we repeat the experiments on attacking the CIFAR10 robust model but replace the normal local models with robust local models, which are adversarially trained DenseNet and ResNet models mentioned in Section \ref{subsec:setupmodels}.\footnote{We did not repeat the experiments with robust MNIST local models because, without worrying about separately training robust local models, we can simply improve the attack performance significantly by tuning the local models during the hybrid attack process (see Table~\ref{tab:hypo_fine_tune} in Section~\ref{sec:loc_model_tune}). The tuning process transforms the normal local models into more robust ones (details in Section~\ref{sec:loc_model_tune}).}

Table~\ref{tab:cifar10_trans_local_model_comp} compares the direct transfer rates for adversarial example candidates found using ensembles of normal and robust models against both types of target models. We see that using robust models in the local ensemble increases the direct transfer rate against the robust model from $10.1\%$ to $40.7\%$ (while reducing the transfer rate against the normal target model). 
We also find that the candidate adversarial examples found using robust local models also provide better starting points for gradient black-box attacks. For example, with the AutoZOOM attack, the mean cost reduction with respect to the baseline mean query (2,632) is significantly improved (from $3.8\%$ to $20.5\%$). We also observe a significant increase of fraction better (percentages of seeds that starting from the local adversarial example is better than starting from the original seed) from $73.1\%$ to $95.5\%$, and a slight increase in the overall success rate of the hybrid attack (from 65.3\% to 68.7\%). When an ensemble of robust local models is used to attack normal target models, however, the attack efficiency degrades significantly, supporting our hypothesis that robust and normal models have different attack surfaces. 

\shortsection{Universal Local Ensemble} 
The results above validate our hypothesis that the different attack surfaces of robust and normal models cause the ineffectiveness against the robust CIFAR10 model in Table~\ref{tab:hypo_loc_adv_sample}. Therefore, to achieve better performance, depending on the target model type, the attacker should selectively choose the local models. However, in practice, attackers may not know if the target model is robustly trained, so cannot predetermine the best local models. We next explore if a universal local model ensemble exists that works well for both normal and robust target models.

To look for the best local ensemble, we tried all 31 different combinations of the 5 local models (3 normal and 2 robust) and measured their corresponding direct transfer rates against both normal and robust target models. 
Figure~\ref{fig:local_ensembles} reports the transfer rates for each local ensemble against both normal and robust target models. For clarity in presentation, we only include results for the five individual models and four representative ensembles in the figure: 
an ensemble of \model{NA}-\model{NB} (selected to represent the mediocre case); \model{Robust-2}, an ensemble of the two robust models (\model{R-DenseNet} and \model{R-ResNet}); \model{Normal-3}, an ensemble of three normal models (\model{NA}, \model{NB}, and \model{NC}); and \model{All-5}, an ensemble of all five local models. These include the ensembles that have the highest or lowest transfer rates to the target models and transfer rates of all other ensembles fit between the reported highest and lowest values. 

None of the ensembles we tested had high direct transfer rates against both normal and robust target models. Ensembles with good performance against robust targets have poor performance against normal targets (e.g., \model{Robust-2} has 37.8\% transfer rate to robust target, but 18.1\% to normal target), and ensembles that have good performance against normal targets are bad against robust targets (e.g., \model{Normal-3} has 65.6\% transfer rate to the normal target, but only 9.4\% to the robust target). Some ensembles are mediocre against both (e.g., \model{NA}-\model{NB}).

One possible reason for the failure of ensembles to apply to both types of target, is that when white-box attacks are applied on the mixed ensembles, the attacks still ``focus'' on the normal models as normal models are easier to attack (i.e., to significantly decrease the loss function). Biasing towards normal models makes the candidate adversarial example less likely to transfer to a robust target model. This conjecture is supported by the observation that although the mixtures of normal and robust models mostly fail against robust target models, they still have reasonable transfer rates to normal target models (e.g., ensemble of 5 local models has 63.5\% transfer rate to normal CIFAR10 target model while only 9.5\% transfer rate to the robust target model). It might be interesting to explore if one can explicitly enforce the attack to focus more on the robust model when attacking the mixture of normal and robust models.

In practice, attackers can dynamically adapt their local ensemble based on observed results, trying different local ensembles against a particular target for the first set of attempts and measuring their transfer rate, and then selecting the one that worked best for future attacks. This simulation process adds overhead and complexity to the attack, but may still be worthwhile when the transfer success rates vary so much for different local ensembles.

For our subsequent experiments on CIFAR10 models, we use the \model{Normal-3} and \model{Robust-2} ensembles as these give the highest transfer rates to normal and robust target models.

\subsection{Local Model Tuning}\label{sec:loc_model_tune}

To test the hypothesis that the labels learned from optimization attacks can be used to tune local models, we measure the impact of tuning on the local models' transfer rate. 

During black-box gradient attacks, there are two different types of input-label pairs generated. One type is produced by adding small magnitudes of random noise to the current image to estimate target model gradients. The other type is generated by perturbing the current image in the direction of estimated gradients. We only use the latter input-label pairs as they contain richer information about the target model boundary since the perturbed image moves towards the decision boundary. These by-products of the black-box attack search can be used to retrain the local models (line~\ref{line:tunemodels} in Algorithm~\ref{alg:hybrid}). The newly generated image and label pairs are added to the original training set to form the new training set, and the local models are fine-tuned on the new training set. As more images are attacked, the training set size can quickly explode. To avoid this, when the size of new training set exceeds a certain threshold $c$, we randomly sample $c$ of the training data and conduct fine-tuning using the sampled training set. For MNIST and CIFAR10, we set the threshold $c$ as the standard training data size (60,000 for MNIST and 50,000 for CIFAR10). At the beginning of hybrid attack, the training set consists of the original seeds available to the attacker with their ground-truth labels (i.e., 1,000 seeds for MNIST and CIFAR10 shown in Section~\ref{subsec:setupattacks}). 


\begin{table}[tb]
\scalebox{1.0}{
\begin{tabular}{ccrr}
\toprule
\multirow{2}{*}{Model} & \multirow{2}{*}{Gradient Attack} & \multicolumn{2}{c}{Transfer Rate (\%)} \\
 &  & {\em Static} & {\em Tuned} \\ \hline
\multirow{2}{*}{MNIST (N, t)} & AutoZOOM & 60.6 & \textbf{64.4} \\
 & NES & 60.6 & \textbf{77.9} \\ \hline
\multirow{2}{*}{MNIST ({\bf R}, u)} & AutoZOOM & 3.4 & \textbf{4.3} \\
 & NES & 3.4 & \textbf{4.5} \\ \hline
\multirow{2}{*}{CIFAR10 (N, t)} & AutoZOOM & \textbf{65.6} & 8.6 \\
 & NES & \textbf{65.6} & 33.4 \\ \hline
\multirow{2}{*}{CIFAR10 ({\bf R}, u)} & AutoZOOM & \textbf{9.4} & 8.8 \\
 & NES & \textbf{9.4} & {9.3} \\ \bottomrule
\end{tabular}
}
\caption{Impact of tuning local models on transfer rates (Baseline + Hypothesis 2): {gradient attacks start from original seed}. Transfer rate is measured on independently sampled test images and is averaged over 5 runs. The results for (N, t) are targeted attacks on normal models; ({\bf R}, u) are untargeted attacks on robust models.} 
\label{tab:hypo_2_validate}
\end{table}

\begin{table*}[tb]
\centering
\begingroup
\setlength{\tabcolsep}{5pt} 

\begin{tabular}{ccrrrrrr}


\toprule
\multirow{2}{*}{Model} & 
\multicolumn{1}{c}{Gradient} & 
\multicolumn{2}{c}{Queries/AE} & 
\multicolumn{2}{c}{Success Rate (\%)} & \multicolumn{2}{c}{Transfer Rate (\%)}
\\ 
& \multicolumn{1}{c}{Attack} & \multicolumn{1}{r}{\em Static} & \multicolumn{1}{r}{\em Tuned} & \multicolumn{1}{r}{\em Static} & \multicolumn{1}{r}{\em Tuned} & \multicolumn{1}{r}{\em Static} & \multicolumn{1}{r}{\em Tuned} \\ \hline

\multirow{2}{*}{MNIST Normal (T)} & AutoZOOM & 282 & \textbf{194} & 98.9 & \textbf{99.5} & 60.6 & \textbf{74.7} \\ 
  & NES & 1,000 & \textbf{671} & 89.2 & \textbf{92.2} & 60.6 & \textbf{76.9} \\ 
 \hline
 \multirow{2}{*}{MNIST Robust (U)} & AutoZOOM & 49,776 & \textbf{42,755} & 7.5 & \textbf{8.6} & 3.4 & \textbf{5.1} \\ 
  & NES & 69,275 & \textbf{51,429} & 5.5 & \textbf{7.3} & 3.4 & \textbf{4.8} \\ 
\hline
\multirow{2}{*}{CIFAR10 Normal (T)} & AutoZOOM & \textbf{276} & 459 & \textbf{98.2} & 96.3 & \textbf{65.6} & 19.7 \\ 
  & NES & \textbf{340} & 427 & \textbf{99.8} & 99.6 & \textbf{65.6} & 40.7 \\ 
 \hline
 \multirow{2}{*}{CIFAR10 Robust (U)} & AutoZOOM & \textbf{2,532} & 2,564 & \textbf{65.3} & 64.9 & {9.4} & \textbf{10.1} \\ 
  & NES & 7,317 & \textbf{7,303} & \textbf{38.0} & {37.6} & 9.4 & \textbf{10.7} \\
\bottomrule
\end{tabular}

\endgroup
\caption{Impact of tuning local models (averaged 5 times). Transfer rate is measured on independently sampled test images.}
\vspace{-1em}
\label{tab:hypo_fine_tune}
\end{table*}







Algorithm~\ref{alg:hybrid} shows the local model being updated after every seed, but considering the computational cost required for tuning, we only update the model periodically. For MNIST, we update the model after every 50 seeds; for CIFAR10, we update after 100 seeds (we were not able to conduct the tuning experiments for the ImageNet models because of the high cost of each attack and of retraining). To check the transferability of the tuned local models, we independently sample 100 unseen images from each of the 10 classes, use the local model ensemble to find candidate adversarial examples, and test the candidate adversarial examples on the black-box target model to measure the transfer rate. 

We first test whether the local model can be fine-tuned by the label by-products of baseline gradient attacks (Baseline attack + H2) by checking the transfer rate of local models before and after the fine-tuning process. We then test whether attack efficiency of hybrid attack can be boosted by fine-tuning local models during the attack process (Baseline attack + H1 + H2) by reporting their average query cost and attack success rate. The first experiment helps us to check applicability of H2 without worrying about possible interactions between H2 with other hypotheses. The second experiment evaluates how much attackers can benefit from fine-tuning the local models in combination with hybrid attacks.

We report the results of the first experiment in Table \ref{tab:hypo_2_validate}. For the MNIST model, we observe increases in the transfer rate of local models by fine-tuning using the byproducts of both attack methods---the transfer rate increases from 60.6\% to 77.9\% for NES, and from 60.6\% to 64.4\% for AutoZOOM. Even against the robust MNIST models, the transfer rate improves from the initial value of 3.4\% to 4.3\% (AutoZOOM) and 4.5\% (NES). However, for CIFAR10 dataset, we observe a significant decrease in transfer rate. For the normal CIFAR10 target model, the original transfer rate is as high as 65.6\%, but with fine-tuning, the transfer rate decrease significantly (decreased to 8.6\% and 33.4\% for AutoZOOM and NES respectively). A similar trend is also observed for the robust CIFAR10 target model. These results suggest that the examples used in the attacks are less useful as training examples for the CIFAR10 model than the original training set.  

Our second experiment, reported in Table~\ref{tab:hypo_fine_tune}, combines the model tuning with the hybrid attack. Through our experiments, we observe that for MNIST models, the transfer rate also increases significantly by fine-tuning the local models. For the MNIST normal models, the (targeted) transfer rate increases from the original 60.6\% to 74.7\% and 76.9\% for AutoZOOM and NES, respectively. The improved transfer rate is also higher than the results reported in first experiment. For the AutoZOOM attack, in the first experiment, the transfer rate can only be improved from 60.6\% to 64.4\% while in the second experiment, it is improved from 60.6\% to 76.9\%. Therefore, there might be some boosting effects by taking local AEs as starting points for gradient attacks. 
For the Madry robust model on MNIST, the low (untargeted) transfer rate improves by a relatively large amount, from the original 3.4\% to 5.1\% for AutoZOOM and 4.8\% for NES (still a low transfer rate, but a 41\% relative improvement over the original local model). The local models become more robust during the fine-tuning process. For example, with the NES attack, the local model attack success rate (attack success is defined as compromising all the local models) decreases significantly from the original 96.6\% to 25.2\%, which indicates the tuned local models are more resistant to the PGD attack. The improvements in transferability, obtained as a free by-product of the gradient attack, also lead to substantial cost reductions for the attack on MNIST, as seen in Table \ref{tab:hypo_fine_tune}. For example, for the AutoZOOM attack on the MNIST normal model, the mean query cost is reduced by 31\%, from 282 to 194 and the attack success rate is also increased slightly, from 98.9\% for static local models to 99.5\% for tuned local models. We observe similar patterns for robust MNIST model and demonstrate that Hypothesis 2 also holds on the MNIST dataset. 

However, for CIFAR10, we still find no benefits from the tuning. Indeed, the transfer rate decreases, reducing both the attack success rate and increasing its mean query cost (Table~\ref{tab:hypo_fine_tune}). We do not have a clear understanding of the reasons the CIFAR10 tuning fails, but speculate it is related to the difficulty of training CIFAR10 models. The results returned from gradient-based attacks are highly similar to a particular seed and may not be diverse enough to train effective local models. This is consistent with Carlini et al.'s findings that MNIST models tend to learn well from outliers (e.g., unnatural images) whereas more realistic datasets like CIFAR10 tend to learn well from more prototypical (e.g., natural) examples~\cite{carlini2018prototypical}. Therefore, fine-tuning CIFAR10 models using label by-products, which are more likely to be outliers, may diminish learning effectiveness. Potential solutions to this problem include tuning the local model with mixture of normal seeds and attack by-products. One may also consider keeping some fraction of model ensembles fixed during the fine-tuning process such that when by-products mislead the tuning process, these fixed models can mitigate the problem. We leave further exploration of this for future work.

\section{Batch Attacks}\label{sec:batch_attack}
    
Section~\ref{sec:hybrid_evaluation} evaluates attacks assuming an attacker wants to attack every seed from some fixed set of initial seeds. In more realistic attack scenarios, each query to the model has some cost or risk to the attacker, and the attacker's goal is to find as many adversarial examples as possible using a limited total number of queries. Carlini et al.\ show that, defenders can identify purposeful queries for adversarial examples based on past queries and therefore, detection risk will increase significantly when many queries are made~\cite{chen2019stateful}. We call these attack scenarios \emph{batch attacks}. To be efficient in these resource-limited settings, attackers should prioritize ``easy-to-attack'' seeds. 

A seed prioritization strategy can easily be incorporated into the hybrid attack algorithm by defining the $\textit{selectSeed}$ function used in step \ref{line:selectseed} in Algorithm \ref{alg:hybrid} to return the most promising seed: 

$$
\argmin_{\mathbf{x} \in \mathbf{X}} \text{\em EstimatedAttackCost}(\mathbf{x}, F).
$$ 

To clearly present the hybrid attack strategy in the batch setting, we present a two-phase strategy: in the first phase, local model information is utilized to find likely-to-transfer
seeds; in the second phase, target model information is used to select candidates for optimization attacks. This split reduces the generality of the attack, but simplifies our presentation and analysis. Since direct transfers have such low cost (that is, one query when they succeed) compared to the optimization attacks, constraining the attack to try all the transfer candidates first does not compromise efficiency.  More advanced attacks might attempt multiple transfer attempts per seed, in which case the decision may be less clear when to switch to an optimization attack. We do not consider such attacks here.


\begin{figure*}[tbh]
        \centering
        \begin{subfigure}[b]{0.49\textwidth}
        \centering
        \includegraphics[width=0.9\textwidth]{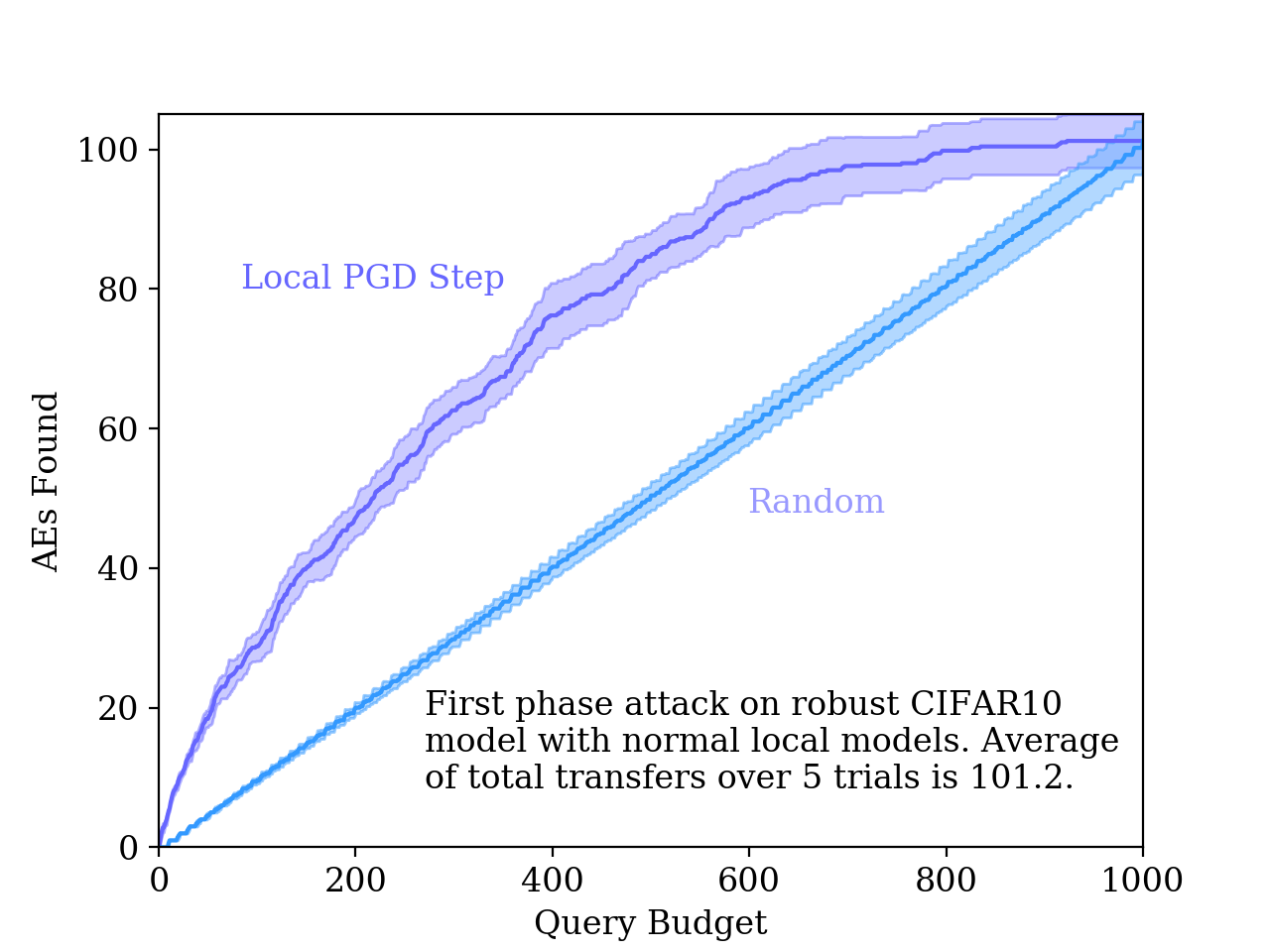}
        \caption[]%
        {Local Normal-3 Ensemble: \model{NA}, \model{NB}, \model{NC}}    
        \label{fig:first_stage_robust_untar_cifar10_NB_d_e}
        \end{subfigure}
        \hfill
        \begin{subfigure}[b]{0.49\textwidth}  
            \centering 
            \includegraphics[width=0.9\textwidth]{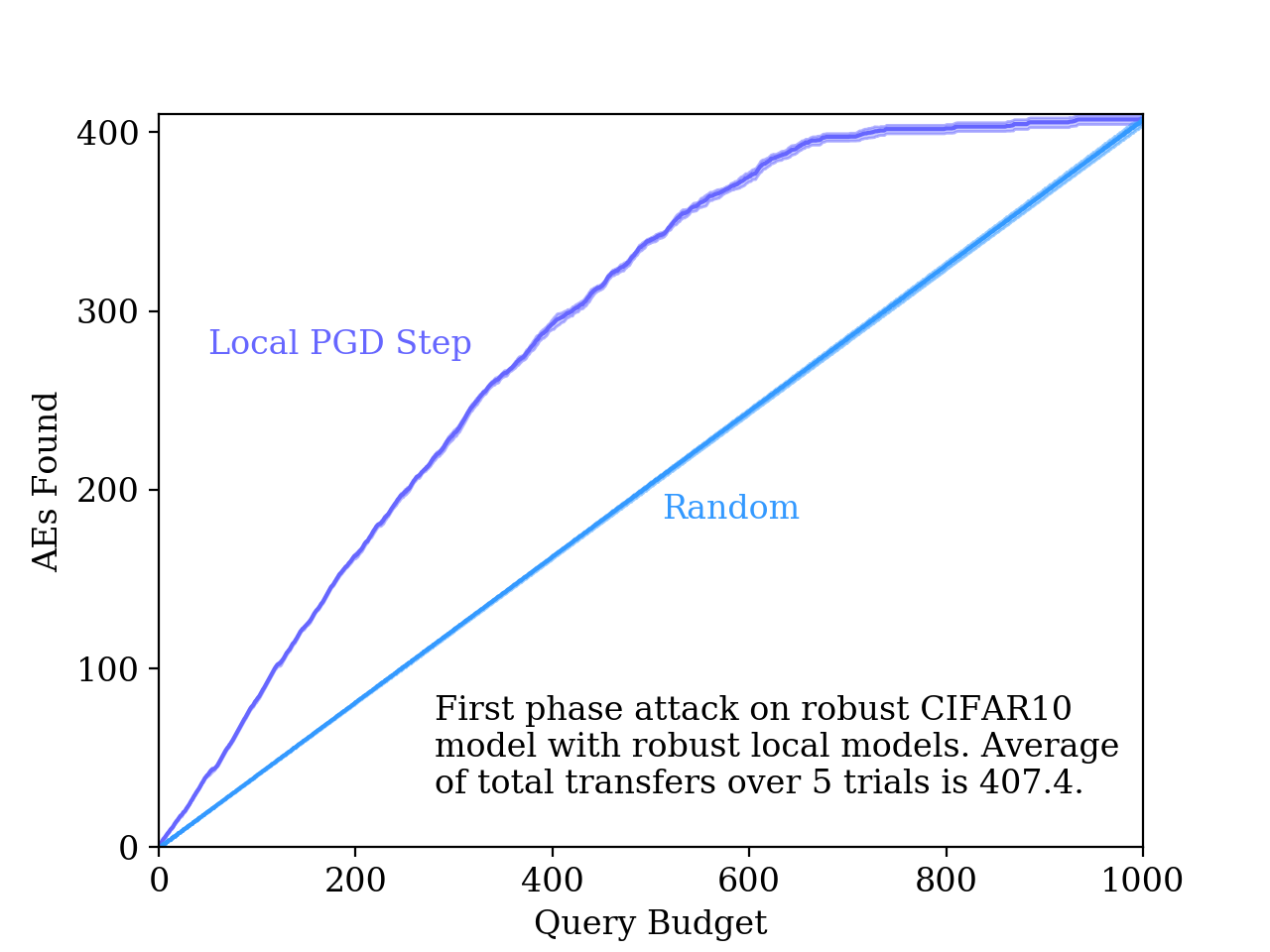}
            \caption[]%
            {Local Robust-2 Ensemble: \model{R-DenseNet}, \model{R-ResNet}}    
            \label{fig:first_stage_robust_untar_cifar10_densenet_resnet}
        \end{subfigure}
        \caption[]
         {First phase (transfer only) attack prioritization (untargeted attack on robust CIFAR10 model, average over 5 runs). Solid line denotes the mean value and shaded area denotes the 95\% confidence interval.} 
        \label{fig:first_stage_robust_untar_cifar10}
    \end{figure*}

\begin{table*}[!tbp]
\centering

\begingroup
\setlength{\tabcolsep}{0.75em}
\renewcommand{\arraystretch}{1.2}

\begin{tabular}{ccrrrr}
\toprule
Local Ensemble & Metric & \multicolumn{1}{c}{First AE} & \multicolumn{1}{c}{Top 1\%} & \multicolumn{1}{c}{Top 2\%} & \multicolumn{1}{c}{Top 5\%} \\ \hline
\multirow{2}{*}{Normal-3} & Local PGD Step & $1.4\pm{0.5}$ & $20.4\pm{2.1}$ & $54.2\pm{5.6}$ & $218.2\pm{28.1}$ \\ 
 & Random & $11.4\pm{0.5}$ & $100.8\pm{4.9}$ & $199.6\pm{9.7}$ & $496.6\pm{24.2}$ \\ \hline

\multirow{2}{*}{Robust-2} & Local PGD Step & $1.0\pm{0.0}$ & $11.8\pm{0.4}$ & $25.6\pm{0.9}$ & $63.8\pm{0.8~~}$ \\ 
& Random & $4.0\pm{0.0}$ & $26.0\pm{0.0}$ & $50.4\pm{0.5}$ & $124.2\pm{1.3~~}$ \\ \bottomrule 
\end{tabular}
\caption[]%
{Impact of prioritization for first phase (robust CIFAR10 Model, average over 5 runs).}
\label{appex_tab:first_stage_comp}
\endgroup
\end{table*}

\subsection{First Phase: Transfer Attack}\label{sec:seed_prior_first_stage}

Since the first phase seeks to find direct transfers, it needs to execute without any information from the target model.  The goal is to order the seeds by likelihood of finding a direct transfer before any query is done to the model. As before, we do assume the attacker has access to pretrained local models, so can use those models both to find candidates for transfer attacks and to prioritize the seeds.

Within the transfer attack phase, we use a prioritization strategy based on the number of PGD-Steps of the local models to predict the transfer likelihood of each image. We explored using other metrics based on local model information such as local model attack loss and local prediction score gap (difference in the prediction confidence of highest and second highest class), but did not find significant differences in the prioritization performance compared to PGD-Step. Hence, we only present results using PGD-Steps here.

\shortsection{Prioritizing based on PGD Steps}
We surmised that the easier it is to find an adversarial example against the local models for a seed, the more likely that seed has a large vulnerability region in the target model. One way to measure this difficult is the number of PGD steps used to find a successful local adversarial example and we prioritize seeds that require less number of PGD steps. To be more specific, we first group images by their number of successfully attacked local models (e.g., $k$ out of $K$ local models), and then prioritize images in each group based on their number of PGD steps used to find the adversarial examples that compromises the $k$ local models. We prioritize adversarial examples that succeed against more of the local models (i.e., larger value of $k$) with the assumption that adversarial examples succeed on more local models tend to have higher chance to transfer to the ``unknown'' target model. Above prioritization strategy is the combination of the metrics of number of successfully compromised local models and PGD steps. We also independently tested the impact of each of the two metrics, and found that the PGD-step based metrics perform better than the number of successfully attacked models, and our current metric of combining the number of PGD steps and the number of successfully attacked models is more stable compared to just using the PGD steps. 

\shortsection{Results}\label{sec:first_stage_eval}
Our prioritization strategy in the first phase sorts images and each seed is queried once to obtain direct transfers. We compare with the baseline of \emph{random} selection of seeds where the attacker queries each seed once in random order to show the advantage of prioritizing seeds based on PGD-Steps. 

Figure~\ref{fig:first_stage_robust_untar_cifar10} shows the results of untargeted attack 
on the Madry robust CIFAR10 model for both normal and robust local model ensembles. Note that first phase attack only checks transferability of the candidate adversarial examples and is independent from the black-box optimization attacks. All results are averaged over five runs.
In all cases, we observe that, checking transferability with prioritized order in the first phase is significantly better than checking the transferability in random order. More quantitative information is given in Table \ref{appex_tab:first_stage_comp}. For the untargeted
attack on robust CIFAR10 model with the three normal local models (\model{NA}, \model{NB}, \model{NC}), when attacker is interested in obtaining $1\%$ of the total 1,000 seeds, checking transferability with prioritized order reduces the cost substantially---with prioritization, it takes 20.4 queries on average, compared to 100.8 with random order. We observed similar patterns for other datasets and models.

    \begin{figure*}[!tbh] 
        \centering
        \begin{subfigure}[b]{0.49\textwidth}
            \centering
            \includegraphics[width=0.9\textwidth]{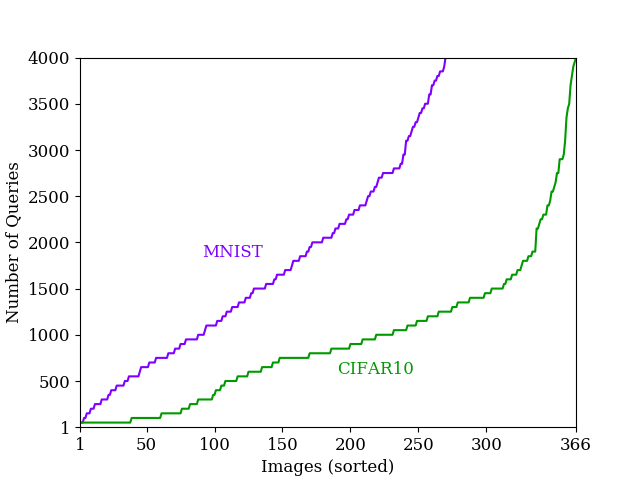}
            \caption[]
            {MNIST and CIFAR10} 
            \label{fig:query_distirbution_mnist_cifar}
        \end{subfigure}
        \hfill
        \begin{subfigure}[b]{0.49\textwidth}  
            \centering
            \includegraphics[width=0.9\textwidth]{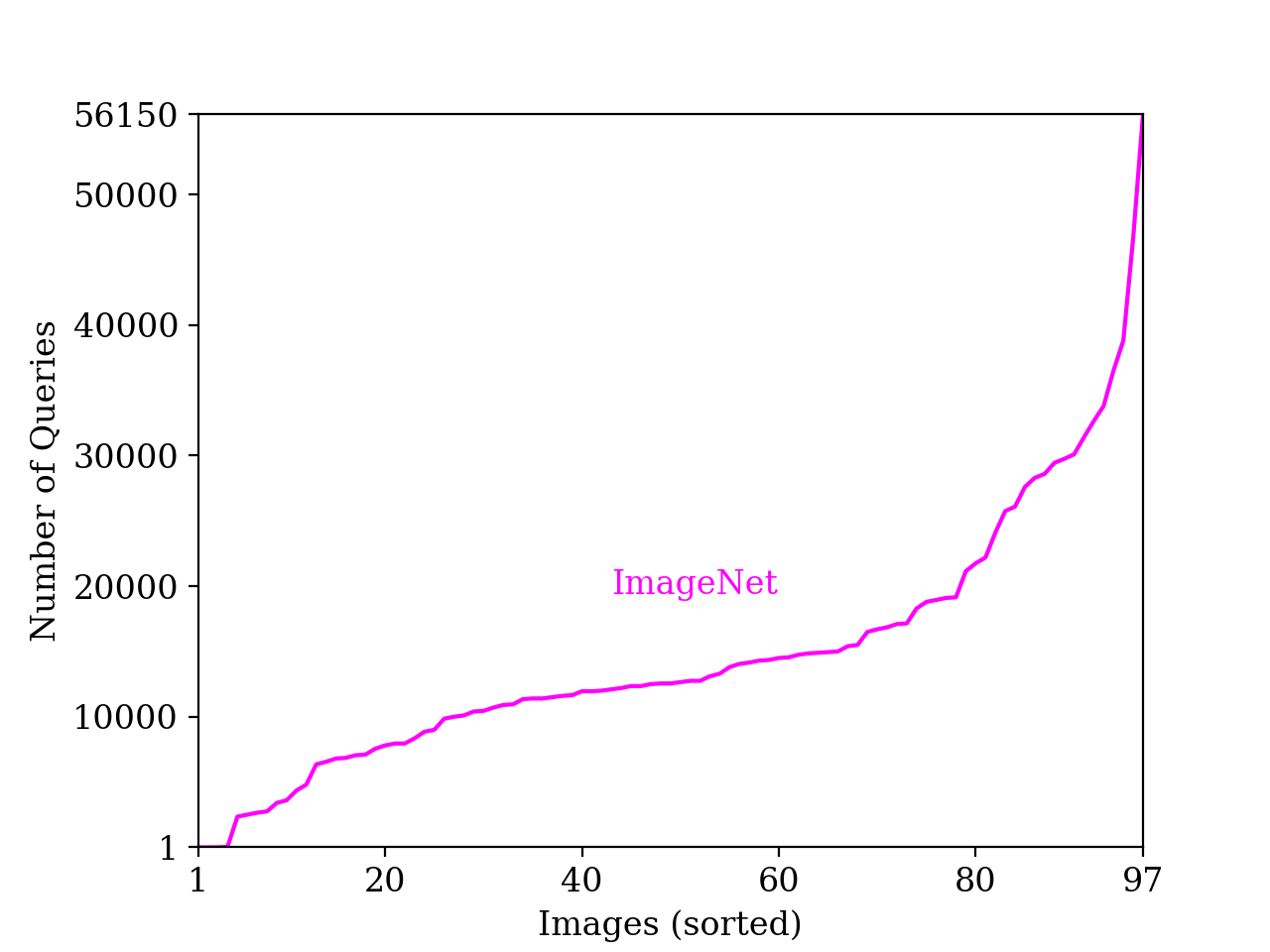}
            \caption[]
            {ImageNet}    
            \label{fig:query_distirbution_imagenet}
        \end{subfigure}
        \caption{Query cost of NES attack on MNIST, CIFAR10 and ImageNet models. We exclude direct transfers (successfully attacked during first phase) and seeds for which no adversarial example was found within the query limit (4000 for MNIST and CIFAR; 10,000 for ImageNet). All the target models are normal models with NES targeted attacks.} 
        \label{fig:query_distirbution}
    \end{figure*}

\subsection{Second Phase: Optimization Attacks}\label{sec:seed_prior_second_stage}
The transfer attack used in the first phase is query efficient, but has low success rate. Hence, when it does not find enough adversarial examples, the attack continues by attempting the optimization attacks on the remaining images. In this section, we show that the cost of the optimization attacks on these images varies substantially, and then evaluate prioritization strategies to identify low-cost seeds. 

\shortsection{Query Cost Variance of Non-transfers}
Figure \ref{fig:query_distirbution} shows the query distributions of non-transferable images for MNIST, CIFAR10 and ImageNet using the NES attack starting from local adversarial examples (similar patterns are observed for the AutoZOOM attack). For ImageNet, when images are sorted by query cost, the top 10\% of 97 images (excluding 3 direct transfers and 0 failed adversarial examples from the original 100 images) only takes on average 1,522 queries while the mean query cost of all 100 images is 14,828. So, an attacker interested in obtaining only 10\% of the total 100 seeds using this prioritization reduces their cost by 90\% compared to targeting seeds randomly. The impact is even higher for CIFAR10 --- the mean query cost for obtaining adversarial examples for 10\% of the seeds remaining after the transfer phase is reduced by nearly 95\% (from 933 to 51) over the random ordering.

\shortsection{Prioritization Strategies}
These results show the potential cost savings from prioritizing seeds in batch attacks, but to be able to exploit the variance we need a way to identify low-cost seeds in advance. We consider two different strategies for estimating the attack cost to implement the estimator for the $\text{\em EstimatedAttackCost}$ function. The first uses same local information as adopted in the first phase: low-cost seeds tend to have lower PGD steps in the local attacks. The drawback of prioritizing all seeds only based on local model information is that local models may not produce useful estimates of the cost of attacking the target model.  Hence, our second prioritization strategy uses information obtained from the single query to the target model that is made for each seed in the first phase.  This query results in obtaining a target model prediction score for each seed, which we use to prioritize the remaining seeds in the second phase. Specifically, we find that low-cost seeds tend to have lower loss function values, defined with respect to the target model. The assumption that an input with a lower loss function value is closer to the attacker's goal is the same assumption that forms the basis of the optimization attacks. 

\begin{figure*}[!tb] 
        \centering
        \begin{subfigure}[b]{0.49\textwidth}  
            \centering 
            \includegraphics[width=0.9\textwidth]{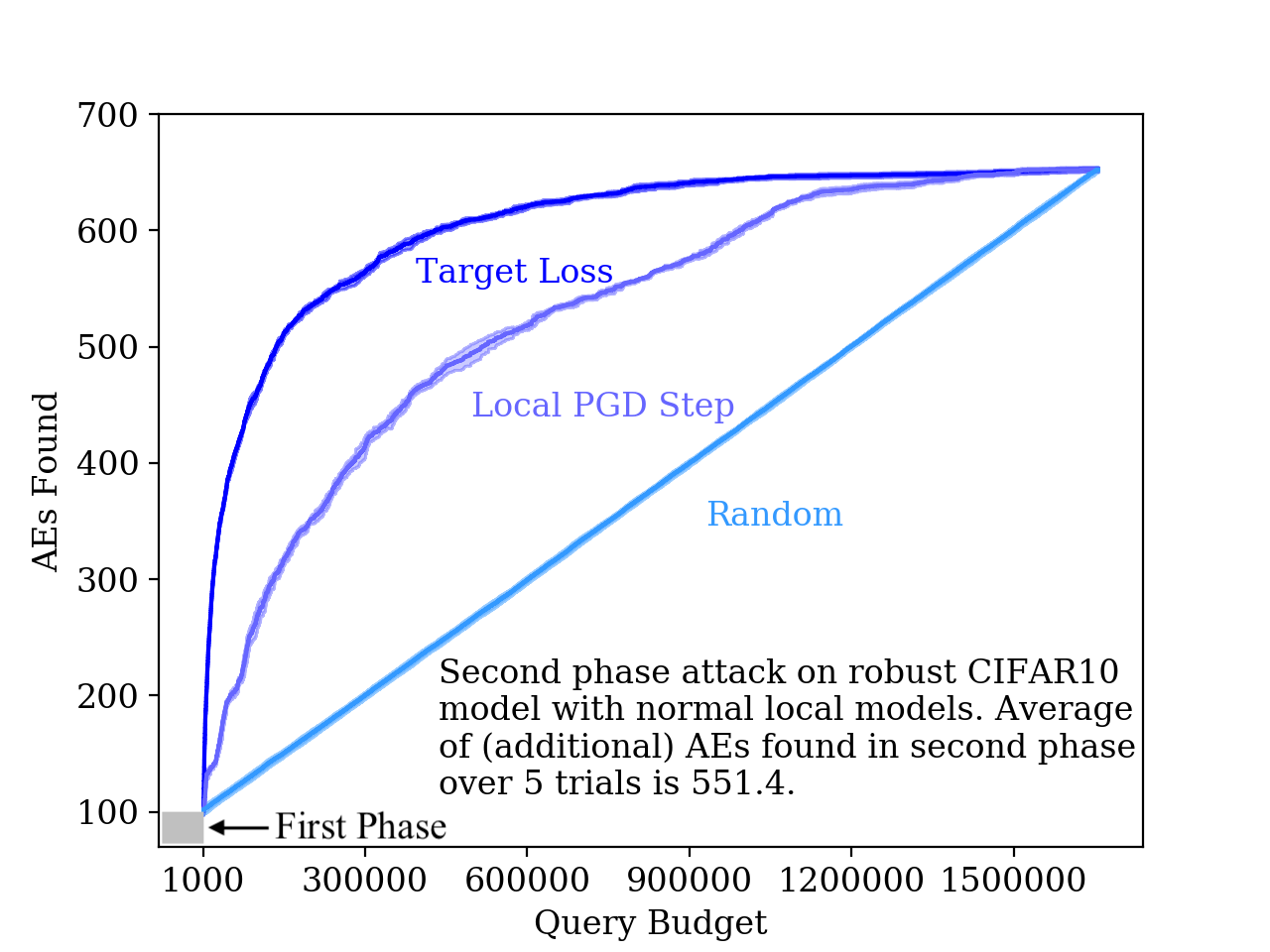}
            \caption[]%
            {Normal-3 Local Ensemble}    
            \label{fig:second_stage_robust_untar_cifar10_NB_d_e}
        \end{subfigure}
        \hfill
        \begin{subfigure}[b]{0.49\textwidth}  
            \centering 
            \includegraphics[width=0.9\textwidth]{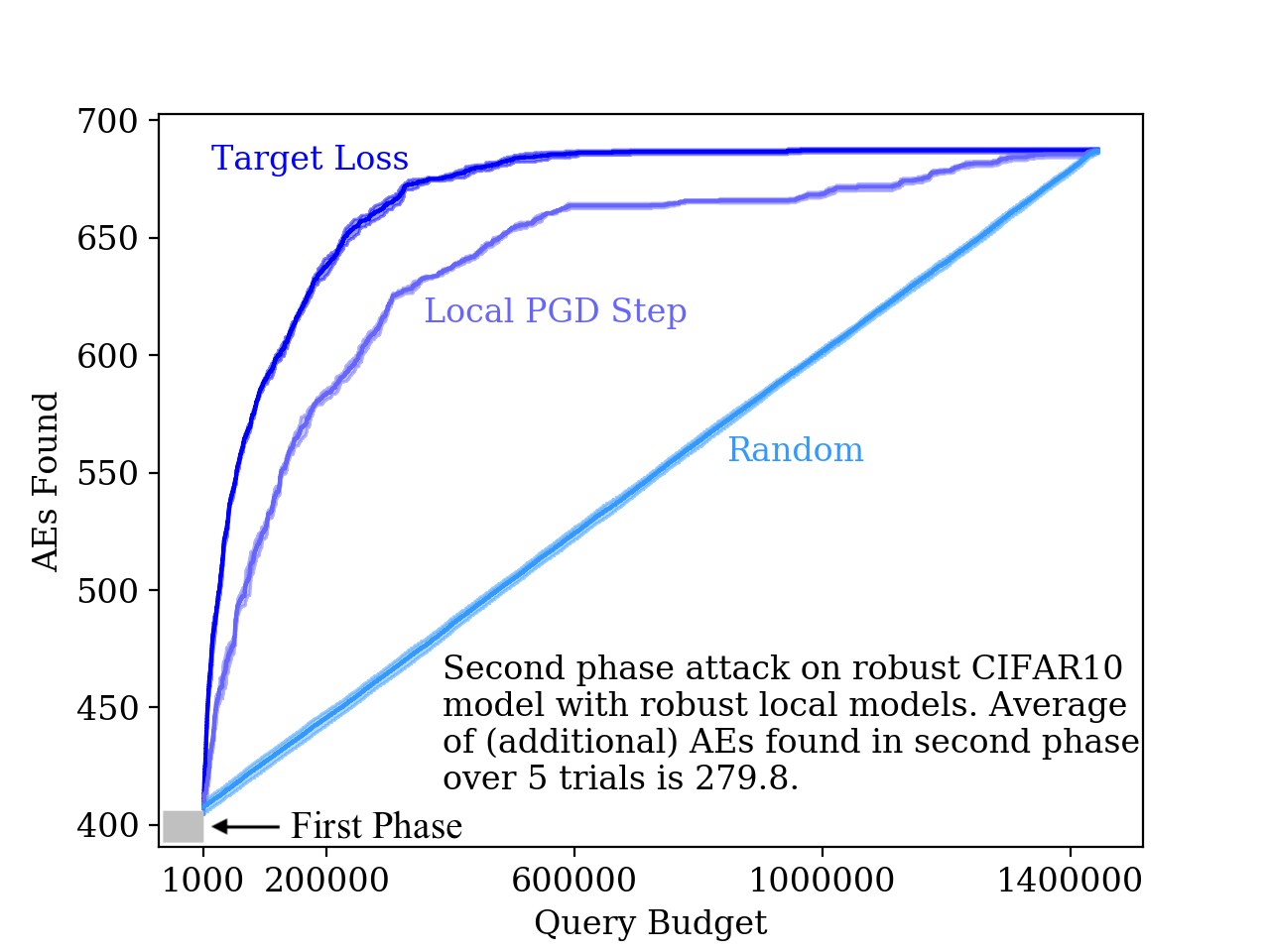}
            \caption[]%
            {Robust-2 Local Ensemble}    
            \label{fig:second_stage_robust_untar_cifar10_densenet_resnet}
        \end{subfigure}
        \caption[]
         {Impact of seed prioritization strategies in the second phase (AutoZOOM untargeted attack on robust CIFAR10 model, average over 5 runs). The x-axis denotes the query budget and the y-axis denotes the number of successful adversarial examples found with the given query budget. The maximum query budget is the sum of the query cost for attacking all seed images (i.e., the total number of queries used to attack all 1000 seeds for CIFAR10 models) --- 1,656,818 for the attack with normal local models, and 1,444,980 for the attack with robust local models. The second phase starts at 1000 queries and the number of direct transfers found because it begins after checking the direct transfers in the first phase.
         } 

        \label{fig:second_stage_robust_untar_cifar10}
    \end{figure*}

 \begin{table*}[tbp]
    \centering
    \begingroup
    \setlength{\tabcolsep}{0.45em}
    \renewcommand{\arraystretch}{1.2}
    \begin{tabular}{
    ccrrrr}
    \toprule
    Local Ensemble & Metric & \multicolumn{1}{c}{Additional 1\%} & \multicolumn{1}{c}{Additional 2\%} & \multicolumn{1}{c}{Additional 5\%} & \multicolumn{1}{c}{Additional 10\%} \\ \hline
    \multirow{3}{*}{\begin{tabular}[c]{@{}c@{}}Normal-3 \end{tabular}} & \begin{tabular}[c]{@{}c}Target Loss \end{tabular} & $1,248\pm93\,~~~~~$ & $1,560\pm147\,~~~$ & $2,739\pm118\,~~~$ & $6,229\pm336\,~~~$  \\ 
     & \begin{tabular}[c]{@{}c}Local PGD Step \end{tabular} & $3,465\pm266\,~~~$ & $4,982\pm274\,~~~$ & $29,203\pm{4,450}$ & $51,962\pm{5,117}$ \\ 
     & \begin{tabular}[c]{@{}c}Random \end{tabular} & $26,336\pm3,486$ & $58,247\pm{3,240}$ & $150,060\pm{3,415}$ & $301,635\pm{6,651}$ \\ \hline
    \multirow{3}{*}{\begin{tabular}[c]{@{}c@{}}Robust-2 \end{tabular}} & \begin{tabular}[c]{@{}c}Target Loss
    \end{tabular} & $2,086\pm37\,~~~~~$ & $3,900\pm604\,~~~$ & $9,882\pm{2,051}$ & $29,435\pm{2,418}$ \\
     & \begin{tabular}[c]{@{}c}Local PGD Step\end{tabular} & $6,009\pm834\,~~~$ & $10,875\pm{1,391}$ & $28,625\pm{3,148}$ & $75,002\pm{5,663}$ \\ 
     & \begin{tabular}[c]{@{}c}Random \end{tabular} & $49,410\pm{1,596}$ & $99,900\pm{3,261}$ & $258,278\pm{2,136}$ & $512,398\pm{6,606}$ \\ \bottomrule
    \end{tabular}
    \caption[]{Impact of different prioritization strategies for optimization attacks (AutoZOOM untargeted attack on robust CIFAR10 model, average over 5 runs). For different models, their number of direct transfers varies---for Normal-3 there are in average 101.2, for Robust-2 there are in average 407.4. We report the number of queries needed to find an additional x\% (10, 20, 50, and 100 out of 1000 total seeds), using the remaining seeds after the first phase. 
    }
    
    \label{appex_tab:second_stage_comp}
    \endgroup
  \end{table*}

Taking a targeted attack as an example, we compute the loss similarly to the loss function used in AutoZOOM~\cite{chen2018AutoZOOM}. For a given input $\mathbf{x}$ and target class $t$, the loss is calculated as 
$$l(\mathbf{x},t) = (\text{max}_{i \neq t} \log f(\mathbf{x})_{i}-\log f(\mathbf{x})_{t})^{+}$$ 
\noindent where $f(\mathbf{x})$ denotes the prediction score distribution of a seed. So, {$f(\mathbf{x})_{i}$} is the model's prediction of the probability that $\mathbf{x}$ is in class $i$. Similarly, for an untargeted attack with original label $y$, the loss is defined as {$l(\mathbf{x},y) = \text{max}(\log f(\mathbf{x})_{y} - \text{max}_{i\neq y} \log f(\mathbf{x})_{i})^{+}$}. Here, the input $\mathbf{x}$ is the candidate starting point for an optimization attack. Thus, for hybrid attacks that start from a local candidate adversarial example, $\mathbf{z}'$, of the original seed $\mathbf{z}$, attack loss is computed with respect to $\mathbf{z}'$ instead of $\mathbf{z}$. For the baseline attack that starts from the original seed $\mathbf{z}$, the loss is computed with respect to $\mathbf{z}$. 

\shortsection{Results}\label{sec:second_stage_eval}
We evaluate the prioritization for the second phase using the same experimental setup as in Section~\ref{sec:first_stage_eval}. We compare the two prioritization strategies (based on local PGD steps and the target model loss) to random ordering of seeds to evaluate their effectiveness in identifying low-cost seeds. The baseline attacks (AutoZOOM and NES, starting from the original seeds) do not have a first phase transfer stage, so we defer the comparison results to next subsection, which shows performance of the combined two-phase attack.

Figure~\ref{fig:second_stage_robust_untar_cifar10}
shows the results for untargeted AutoZOOM attacks on the robust CIFAR10 model for local ensembles of both normal and robust models (results for the NES attack are not shown, but exhibit similar patterns). Using the target loss information estimates the attack cost better than the local PGD step ordering, while both prioritization strategies achieve much better performance than the random ordering. Table \ref{appex_tab:second_stage_comp} summarizes the results. For example, for the untargeted AutoZOOM attack on robust CIFAR10 model with the Normal-3 local ensemble, an attacker who wants to obtain ten new adversarial examples (in addition to the 101 direct transfers found in the first phase) can find them using on average 1,248 queries using target model loss in the second phase, compared to 3,465 queries when using only local ensemble information, and 26,336 without any prioritization.

     \begin{figure*}[tb]
        \centering
        \begin{subfigure}[b]{0.49\textwidth}
            \centering
            \includegraphics[width=0.9\textwidth]{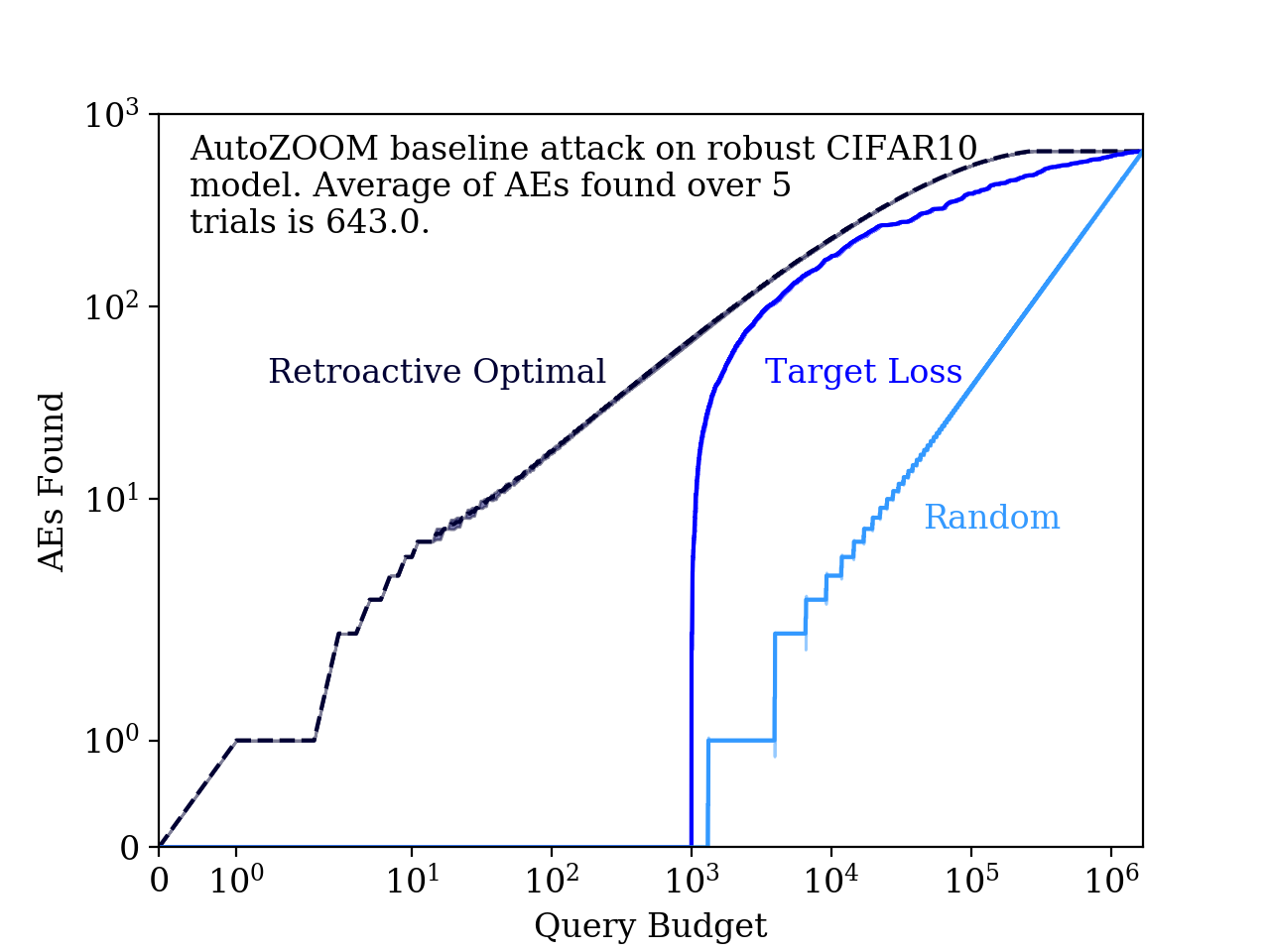}
            \caption[]
            {Target: Robust CIFAR10 Model} 
            \label{fig:h4_test_autozoom_main_cifar10}
        \end{subfigure}
        \begin{subfigure}[b]{0.49\textwidth}  
            \centering 
            \includegraphics[width=0.9\textwidth]{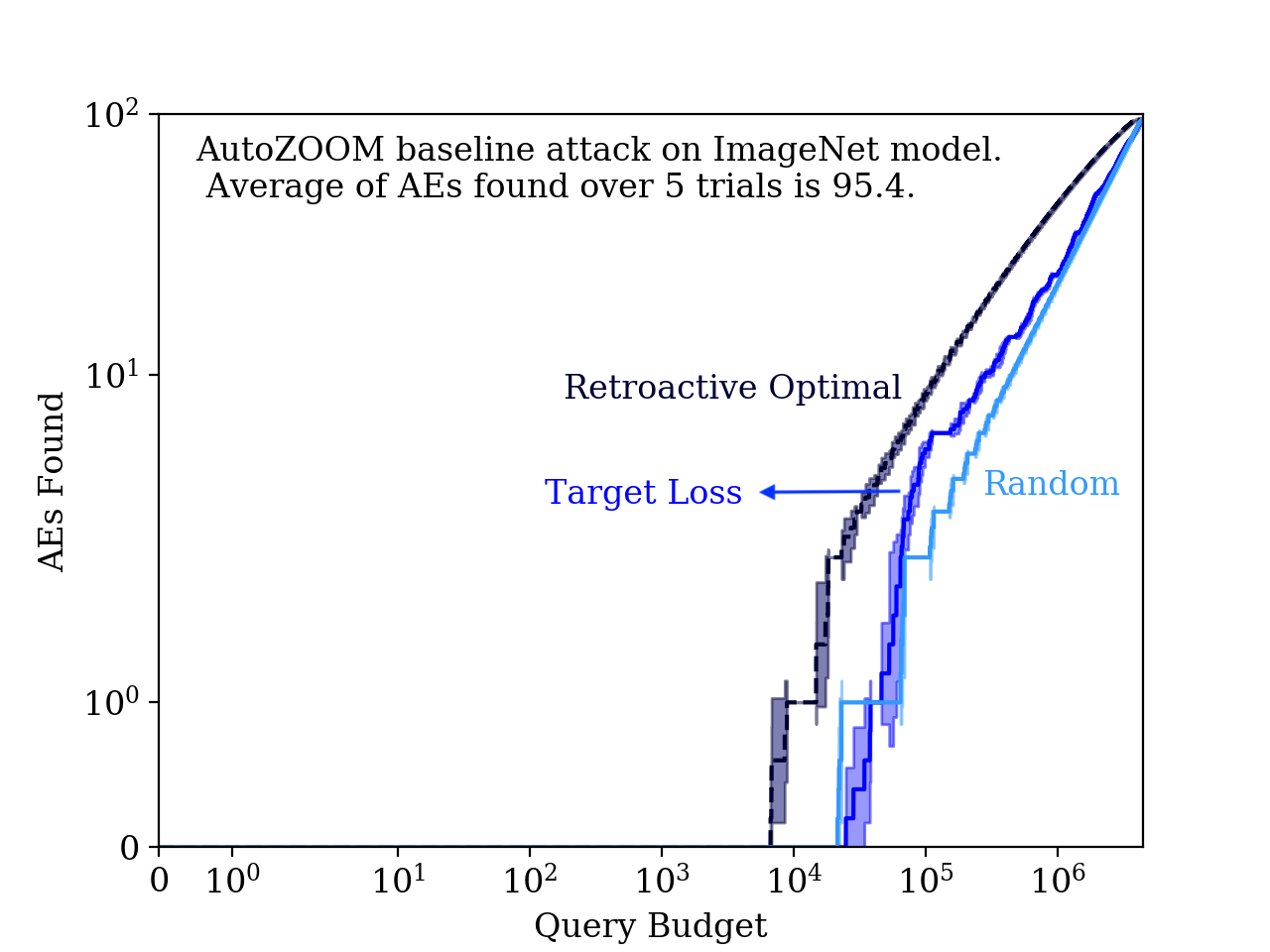}
            \caption[]%
            {Target: Standard ImageNet Model}    
            \label{fig:h4_test_autozoom_main_imagenet}
        \end{subfigure}
        \caption[]
        {Comparison of the target loss value based seed prioritization strategy to retroactive optimal and random search strategies (AutoZOOM baseline untargeted attack on robust CIFAR10 model and targeted attack on standard ImageNet model, averaged over 5 runs). Solid line denotes mean value and shaded area denotes the 95\% confidence interval. Maximum query budget is 1,699,998 for robust CIFAR10 model, 4,393,314 for ImageNet.} 
        \label{fig:h4_test_autozoom_main}
    \end{figure*}

    \begin{table*}[tb]
    \centering
    \renewcommand{\arraystretch}{1.2}
    \begin{tabular}{ccrrrr}
    \toprule
    Target Model & Prioritization Method & \multicolumn{1}{c}{Top 1\%} & \multicolumn{1}{c}{Top 2\%} & \multicolumn{1}{c}{Top 5\%} & \multicolumn{1}{c}{Top 10\%} \\ \hline
    \multirow{3}{*}{\begin{tabular}[c]{@{}c@{}c@{}}Robust\\ CIFAR10 \\ (1,000 Seeds)\end{tabular}} & Retroactive Optimal & $34.0\pm2.0\,~~~~$ & $119.2\pm4.8\,~~~~$ & $580.8\pm35.0\,~~$ & $2,002\pm69\,~~~~~~~$ \\
     & Target Loss & $1,070\pm13\,~~~~~$ & $1,170\pm16\,~~~~~$ & $1,765\pm12\,~~~~~$ & $3,502\pm85\,~~~~~~~$ \\ 
     & Random & $25,005\pm108\,~~~$ & $51,325\pm221\,~~~$ & $130,284\pm561\,~~~$ & $261,883\pm{1,128~~}$ \\ \hline
    \multirow{3}{*}{\begin{tabular}[c]{@{}c@{}c@{}}Standard\\ ImageNet\\(100 Seeds)\end{tabular}} & Retroactive Optimal & $7,492\pm{1,078}$ & $16,590\pm{1,755}$ & $49,255\pm{4,945}$ & $114,832\pm{7,430~~}$ \\
     & Target Loss & $32,490\pm{5,857}$ & $58,665\pm{8,268}$ & $89,541\pm{8,459}$ & $257,594\pm{13,738}$ \\ 
     & Random & $22,178\pm705~~~\,$ & $66,532\pm{2,114}$ & $199,595\pm{6,341~\,}$ & $421,365\pm{13,387}$ \\ \bottomrule
    \end{tabular}

    \caption{Comparison of the target loss value based search to retroactive optimal and random search (AutoZOOM baseline untargeted attack on robust CIFAR10 model, targeted attack on standard ImageNet model, averaged over 5 runs).  The ``Top $x$\%'' columns give the total number of queries needed to find adversarial examples for $x$\% of the total seeds. 
    }
    \label{appex_tab:baseline_prior}
    \end{table*}
    
    \begin{figure*}[!tb]
        \centering
        \begin{subfigure}[b]{0.49\textwidth}
            \centering
            \includegraphics[width=0.9\textwidth]{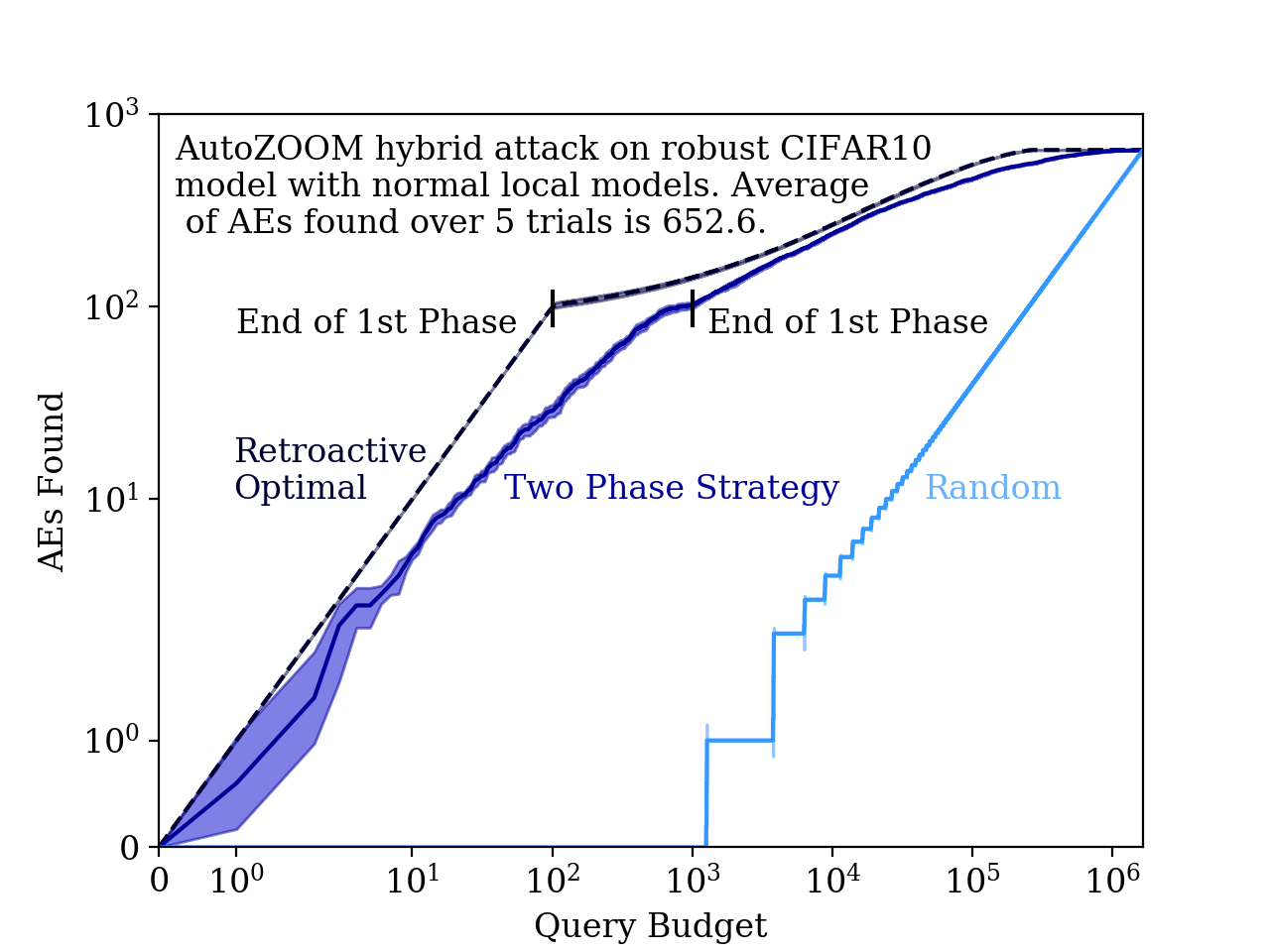}
            \caption[]
            {Target: Robust CIFAR10 Model, Local Ensemble: Normal-3} 
            \label{fig:hybrid_batch_autozoom_main_cifar10}
        \end{subfigure}
        \begin{subfigure}[b]{0.49\textwidth}  
            \centering 
            \includegraphics[width=0.9\textwidth]{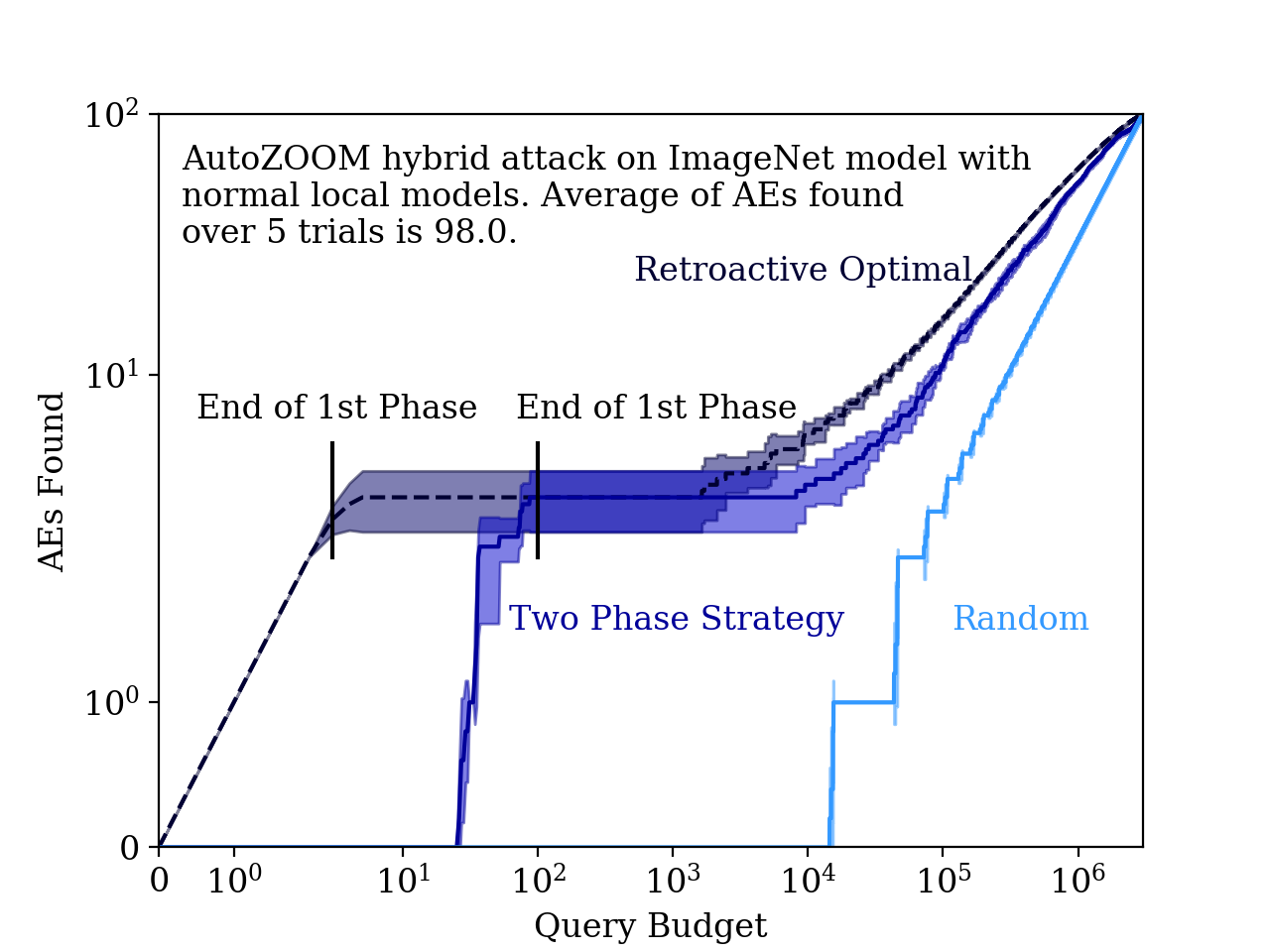}
            \caption[]%
            {Target: Standard ImageNet Model}    
            \label{fig:hybrid_batch_autozoom_main_imagenet}
        \end{subfigure}
        \caption[]
        {Comparison of the two-phase seed prioritization strategy to retroactive optimal and random search strategies (AutoZOOM-based hybrid attack on robust CIFAR10 model and standard ImageNet model, average over 5 runs). Solid line denotes mean value and shaded area denotes the 95\% confidence interval. Maximum query budget of attack against robust CIFAR10 model is 1,656,818 and attack against ImageNet models is 3,029,844.} 
        \label{fig:hybrid_batch_autozoom_main}
    \end{figure*}
    
\subsection{Overall Attack Comparison}\label{sec:prior_metric_eval}
To further validate effectiveness of the seed prioritized two-phase strategy, we evaluate the full attack combining both phases. Based on our analysis in the previous subsections, we use the best prioritization strategies for each phase: PGD-Step in the first phase and target loss value in the second phase. For the baseline attack, we simply adopt the target loss value to prioritize seeds. We evaluate the effectiveness in comparison with two degenerate strategies:
\squishlist
\item \emph{retroactive optimal} --- this strategy is not realizable, but provides an upper bound for the seed prioritization. It assumes the attackers have prior knowledge of the true rank of each seed. That is, we assume a \emph{selectSeed} function that always returns the best remaining seed. 

\item \emph{random} --- 
the attacker selects candidate seeds in a random order and conducts optimization attacks exhaustively (until either success or the query limit is reached) on each seed before trying the next one. This represents traditional black-box attacks that just attack every seed. \squishend




Here, we only present results of AutoZOOM attack on robust CIFAR10 model with normal local models and AutoZOOM attack on ImageNet models. The attack on robust CIFAR10 is the setting where the performance gain for the hybrid attack is least significant compared to other models (see Table \ref{tab:hypo_loc_adv_sample}), so this represents the most challenging scenario for our attack. In the ImageNet setting, the performance of the target loss based prioritization is not a significant improvement over random scheduling, so this represents the worst case for target loss prioritization for the baseline attack. Results of the two black-box attacks on all the other datasets and different combinations of target models and local models (only for the CIFAR10 dataset) show similar patterns.

The results of seed prioritization on baseline attacks are shown in Figure~\ref{fig:h4_test_autozoom_main} and Table~\ref{appex_tab:baseline_prior}. For attacks on the robust CIFAR10 model, performance of the target loss strategy is much better than the random scheduling strategy. For example, in order to obtain 1\% of the total 1,000 seeds, the target loss prioritization strategy costs 1,070 queries on average, while the random strategy consumes on average 25,005 queries, which is a 96\% query savings. The retroactive optimal strategy is very effective in this case and significantly outperforms other strategies by only taking 34 queries. Against the ImageNet model, however, the target loss based strategy offers little improvement over random scheduling (Figure~\ref{fig:h4_test_autozoom_main_imagenet}). In contrast, performance of the two-phase strategy is still significantly better than random ordering. 

 \begin{table*}[tb]
    \centering
    \begingroup
    \setlength{\tabcolsep}{0.35em}
    \renewcommand{\arraystretch}{1.2}
    \scalebox{1.0}{
    \begin{tabular}{ccrrrr}
    \toprule
    {\begin{tabular}[c]{@{}c@{}}Target Model\end{tabular}} & {\begin{tabular}[c]{@{}c@{}}Prioritization Method\end{tabular}} & \multicolumn{1}{r}{Top 1\%} & \multicolumn{1}{r}{Top 2\%} & \multicolumn{1}{c}{~Top 5\%} & \multicolumn{1}{c}{~Top 10\%} \\ \midrule
    \multirow{3}{*}{\begin{tabular}[c]{@{}c@{}c@{}}Robust \\ CIFAR10 \\ (1000 Seeds)\end{tabular}} & {\begin{tabular}[c]{@{}c@{}} Retroactive Optimal\end{tabular}} & $10.0\pm0.0~$ & $20.0\pm0.0~$ & $50.0\pm0.0\,~~~~~~$ & $107.8\pm17.4\,~~~~$  \\
     & Two-Phase Strategy & $20.4\pm2.1~$ & $54.2\pm5.6~$ & $218.2\pm28.2\,~~~~$ & $826.2\pm226.6\,~~$  \\
     & Random & $24,054\pm132$ & $49,372\pm270$ & $125,327\pm686\,~~~~~$ & $251,917\pm137~~~~$ \\ \hline
    \multirow{3}{*}{\begin{tabular}[c]{@{}c@{}c@{}}Standard\\ ImageNet \\ (100 Seeds)\end{tabular}} & {\begin{tabular}[c]{@{}c@{}} Retroactive Optimal\end{tabular}} & $1.0\pm0.0~$ & $2.0\pm0.0~$ & $3,992\pm3,614~~$ & $34,949\pm3,742~~$  \\
     & Two-Phase Strategy& $28.0\pm2.0~$ & $38.6\pm7.5~$ & $18,351\pm13,175$ & $78,844\pm{11,837}$ \\
     & Random & $15,046\pm423$ & $45,136\pm1,270$ & $135,406\pm3,811~~$ & $285,855\pm8045\,~~~$ \\
     \bottomrule
    \end{tabular}
    }
    \endgroup
    \caption{Comparison of the two-phase search to retroactive optimal and random search (AutoZOOM-based hybrid attack on robust CIFAR10 model and standard ImageNet model, average over 5 runs).
    }
    \label{tab:schedule_metric_choice}
    \end{table*} 
We speculate that the difference in the performance of target loss strategy (for baseline attack) and two-phase strategy (for hybrid attack) on ImageNet is because the baseline attack starts from the original seeds, which are natural images and ImageNet models tend to overfit to these natural images. Therefore, the target loss value computed with respect to these images is less helpful in predicting their actual attack cost, which leads to poor prioritization performance. In contrast, the hybrid attack starts from local adversarial examples, which deviate from the natural distribution so ImageNet models are less likely to overfit to these images. Thus, the target loss 
is better correlated with the true attack cost and the prioritization performance is also improved.

Figure~\ref{fig:hybrid_batch_autozoom_main} shows the results for the full two-phase strategy. The seed prioritized two-phase strategy approaches the performance of the (unrealizable) retroactive optimal strategy and substantially outperforms random scheduling. Table~\ref{tab:schedule_metric_choice} shows the number of queries needed using each prioritization method to successfully attack 1\%, 2\%, 5\% and 10\% of the total candidate seeds (1000 images for CIFAR10 and 100 images for ImageNet). For the robust CIFAR10 target model, obtaining 10 new adversarial examples (1\%), costs 20.4 queries on average using our two-phase strategy (not far off the 10 required by the unrealizable retroactive optimal, which takes only a single query for each since it can always find the direct transfer), while random ordering takes 24,054 queries. For ImageNet, the cost of obtaining the first new adversarial example (1\%) using our two-phase strategy is 28 queries compared to over 15,000 with random prioritization.

\section{Conclusion}\label{sec:conclusion}
Our results improve our understanding of black-box attacks against machine learning classifiers and show how efficiently an attacker may be able to successfully attack even robust target models. We propose a hybrid attack strategy, which combines recent transfer-based and optimization-based attacks. Across multiple datasets, our hybrid attack strategy dramatically improves state-of-the-art results in terms of the average query cost, and hence provides more accurate estimation of cost of black-box adversaries. We further consider a more practical attack setting, where the attacker has limited resources and aims to find many adversarial examples with a fixed number of queries. We show that a simple seed prioritization strategy can dramatically improve the overall efficiency of hybrid attacks.

\section*{Availability}
Implementations and data for reproducing our results 
are available at {\url{https://github.com/suyeecav/Hybrid-Attack}}.

\section*{Acknowledgements}
This work was supported by grants from the National Science Foundation (\#1619098, \#1804603, and \#1850479) and research awards from Baidu and Intel, and cloud computing grants from Amazon. 


\bibliographystyle{plain}
\bibliography{bibliography} 

\clearpage

\end{document}